\documentclass[twocolumn, aps, superscriptaddress, prb, showpacs,longbibliography,floatfix]{revtex4-1}
\usepackage{dcolumn}
\usepackage{graphicx}
\usepackage{amsmath,amssymb}
\usepackage{bm}
\usepackage{color}
\usepackage{url}
\usepackage{pict2e,hyperref}

\usepackage{here}

\definecolor{myorange}{rgb}{0.7,0.5,0.0}
\definecolor{mygreen}{rgb}{0.0,0.7,0.0}
\definecolor{purple}{rgb}{0.75,0.0,1.0}

\begin{document}


\title {General corner charge formula in two-dimensional $C_n$-symmetric higher-order topological insulators}
\author {Ryo Takahashi}
\affiliation{
Department of Physics, Tokyo Institute of Technology, 2-12-1 Ookayama, Meguro-ku, Tokyo 152-8551, Japan\\
}
\author {Tiantian Zhang}
\affiliation{
Department of Physics, Tokyo Institute of Technology, 2-12-1 Ookayama, Meguro-ku, Tokyo 152-8551, Japan\\
}
\affiliation{
TIES, Tokyo Institute of Technology, 2-12-1 Ookayama, Meguro-ku, Tokyo 152-8551, Japan\\
}
\author {Shuichi Murakami}
\affiliation{
Department of Physics, Tokyo Institute of Technology, 2-12-1 Ookayama, Meguro-ku, Tokyo 152-8551, Japan\\
}
\affiliation{
TIES, Tokyo Institute of Technology, 2-12-1 Ookayama, Meguro-ku, Tokyo 152-8551, Japan\\
}

\date{\today}

\begin{abstract}
In this paper, we derive a general formula for the quantized fractional corner charge in two-dimensional $C_n$-symmetric higher-order topological insulators. We assume that the electronic states can be described by the Wannier functions and that the edges are charge neutral, but we do not assume vanishing bulk electric polarization. We expand the scope of the corner charge formula obtained in previous works by considering more general surface conditions, such as surfaces with higher Miller index and surfaces with surface reconstruction. Our theory is applicable even when the electronic states are largely modulated near system boundaries. It also applies to insulators with non-vanishing bulk polarization, and we find that in such cases the value of the corner charge depends on the surface termination even for the same bulk crystal with $C_3$ or $C_4$ symmetry, via a difference in the Wyckoff position of the center of the $C_n$-symmetric crystal. 
\end{abstract}

\maketitle

\section{Introduction}
Recent studies have revealed that some topological crystalline insulators exhibit higher-order bulk boundary correspondence, which are known as higher-order topological insulators
\cite{
PhysRevLett.111.047006,
PhysRevB.89.224503,
PhysRevB.95.165443,
fang2019new,
PhysRevLett.119.246401,
PhysRevB.97.155305,
PhysRevB.97.241402,
PhysRevB.97.205135,
PhysRevB.97.205136,
schindler2018higherTI,
schindler2018higher,
PhysRevB.98.081110,
PhysRevX.8.031070,
PhysRevB.98.205129,
wieder2018axion,
PhysRevB.98.245102,
PhysRevX.9.011012,
PhysRevB.99.235125,
PhysRevB.100.205126,
PhysRevB.100.235302,
PhysRevB.101.115120,
PhysRevResearch.2.013300,
PhysRevResearch.2.043274,
Benalcazar61,
PhysRevLett.119.246402,
PhysRevB.96.245115,
serra2018observation,
PhysRevLett.120.026801,
peterson2018quantized,
PhysRevB.98.045125,
imhof2018topolectrical,
PhysRevB.99.245151,
PhysRevLett.122.086804,
PhysRevResearch.2.012009,
doi:10.7566/JPSJ.88.104703,
PhysRevResearch.1.033074,
PhysRevB.101.241109,
PhysRevB.101.115140, 
peterson2020fractional,
PhysRevResearch.2.043131,
PhysRevB.102.165120
}. 
For example, three-dimensional second-order topological insulators with inversion symmetry (and time-reversal symmetry) have protected anomalous gapless mode along the hinges
\cite{
PhysRevB.97.205136,
PhysRevB.98.081110,
PhysRevX.8.031070,
PhysRevB.98.205129,
PhysRevB.98.245102,
PhysRevB.101.115120,
PhysRevResearch.2.013300,
PhysRevResearch.2.043274
}, 
and two-dimensional second-order topological insulators with rotation symmetry have protected quantized corner charges
\cite{
Benalcazar61,
PhysRevLett.119.246402,
PhysRevB.96.245115,
serra2018observation,
PhysRevLett.120.026801,
peterson2018quantized,
PhysRevB.98.045125,
imhof2018topolectrical,
PhysRevB.99.245151,
PhysRevLett.122.086804,
PhysRevResearch.2.012009,
doi:10.7566/JPSJ.88.104703,
PhysRevResearch.1.033074,
PhysRevB.101.115140, 
PhysRevB.101.241109,
peterson2020fractional,
kooi2021bulk,
PhysRevResearch.2.043131,
PhysRevB.102.165120
}. Such fractionally quantized corner charges are generalizations of the quantized surface charge caused by the quantized bulk electric polarization\cite{PhysRevB.48.4442}.

As discussed in the previous works, the quantized fractional corner charge in $C_n$-symmetric crystalline insulators are associated with rotational eigenvalues of the bulk wavefunctions\cite{PhysRevB.99.245151,PhysRevResearch.1.033074,kooi2021bulk,PhysRevResearch.2.043131}; nonetheless, they are limited to special cases. 
For example, in $C_6$-symmetric systems, Ref.~[\onlinecite{PhysRevB.99.245151}] considers the case with hexagonal unit cells that cover the whole crystal, while in Ref.~[\onlinecite{PhysRevResearch.2.043131}] systems covered by triangular building blocks are considered, and so the formulas of the corner charge for these two cases are different. 
In Ref.~[\onlinecite{PhysRevB.102.165120}], although a general formula of the quantized corner charge is discussed, the authors assume that the surface is flat and the electronic charge distribution is the same between the finite $C_6$-symmetric system and an infinite system. 
Most importantly, the formula in Ref.~[\onlinecite{PhysRevB.102.165120}] treats the contributions of electrons and ions to corner charges on an equal footing. 

Here, a question arises whether the formula for the corner charge in previous works \cite{PhysRevB.99.245151,PhysRevResearch.1.033074,kooi2021bulk,PhysRevResearch.2.043131} holds for any insulating $C_n$ symmetric systems; this question is important for application to real materials, which may have deformations of nuclei positions and of electronic states near the boundaries.  
In the previous works, the corner-charge formula was derived for systems in which nuclei positions and electronic states are perfectly periodic up to the boundaries, and this result was complemented by an argument that surface decoration will leave
the corner charge unaffected. 
Nevertheless, the surface-decoration argument has not been addressed in detail, and  it is not clear whether the resulting corner-charge formula holds in 
general $C_n$-symmetric systems, including those where the nuclei positions and electronic states may be modulated near the boundary.

In the present paper we derive a general corner-charge formula on general grounds with minimal assumptions.
Our study shows that even when the positions of the nuclei and the electronic states are largely modulated near the boundaries, the corner charge is quantized to fractional values as long as the system is $C_n$-symmetric and the edge charge density is zero. One of the merits of our method is that it naturally includes  surfaces with higher Miller indices and surfaces with surface reconstructions.
Our theory also includes insulators with non-vanishing bulk polarization but without edge charges. In particular, in such cases with $C_3$ or $C_4$ symmetries, we find that the value of the corner charge depends on the Wyckoff position of the center of the $C_n$-symmetric crystal. It
is in contrast with previous works, where the corner-charge formula assumes zero polarization.

This paper is organized as follows. In Sec.~I\hspace{-.1em}I, we derive the general corner charge formula and clarified the assumptions required for the quantization of the corner charge. We show that as long as the edge is gapped and charge neutral in $C_n$-symmetric finite system, the corner charge is fractionally quantized modulo $|e|/n$. We also discuss the relation of the general formula and the bulk rotational eigenvalues. In Sec.~I\hspace{-.1em}I\hspace{-.1em}I, we consider systems with nonzero electric polarization, which is beyond the scope of the previous corner charge formula. 
In Sec.~I\hspace{-.1em}V, we consider more general surface conditions, such as general edge orientations and surface reconstructions. 
Conclusion and discussion are given in Sec.~V.

\section{Corner charge formula}

In this section, we construct a general theory to calculate the corner charges in an insulator with $C_n$-rotation symmetry, based on the notion of filling anomaly. 
We derive its fractional quantization under the following minimal assumptions. 
We assume that the system is  topologically trivial in
the sense that it is adiabatically connected to an atomic limit \cite{
PhysRevB.99.245151,
PhysRevB.102.165120}.
It assures that the system has no topological edge states in the gap.
  
To explain another important assumption, we show the definition of the filling anomaly adopted in this paper:
\begin{itemize}\item In some bulk insulators, charge neutrality is incompatible with existence of a gap at the Fermi energy for the whole system including the boundries; in such cases, by adding or subtracting a few electrons from charge neutrality, the Fermi energy is shifted and the system becomes gapped including the boundaries. This deviation of the electron number from charge neutrality is called 
filling anomaly in this paper. 
\end{itemize}
Based on this definition of the filling anomaly, we assume that the finite-sized $C_n$-symmetric crystal considered is gapped. These two assumptions, namely topological triviality and existence of the gap, 
guarantee that the occupied states are described in term of localized orbitals. 
We note that the gap in this finite crystal should survive in the large system size limit.
For example, in a hexagonal-shaped crystal with $C_6$ symmetry, suppose the edge and bulk are gapped, and 
the corners support one in-gap bound state per each corner, with the Fermi energy at these in-gap corner 
states. Then a tiny hybridization gap $O(e^{-L/\lambda})$ appears within the sixfold multiplet of the corner states,
where $L$ is the distance between neighboring corners and $\lambda$ is a penetration depth of the corner states
along the edges. In this case, the occupied states at the corners cannot be described as localized orbitals,
but are extended over the system size, and it violates our assumptions. Thus, these assumptions justify the description of occupied electronic 
states as localized orbitals.

In the previous works\cite{
PhysRevB.99.245151,
PhysRevResearch.1.033074,
kooi2021bulk,
PhysRevResearch.2.043131,
PhysRevB.102.165120
}, the formula of the quantized corner charge is derived for cases with the electronic charge distribution being perfectly periodic even near the system boundaries. It is then argued that the quantization will remain even with surface decorations onto the system while preserving the $C_n$ symmetry and the bulk gap. 
Nonetheless, details of surface-decoration argument have not been addressed, and it is not clear in which kinds of systems the corner charge remains quantized to be the value 
determined by bulk electronic states. In real materials, electronic states are determined as solutions of the Schr\"{o}dinger equation, and 
nuclei positions are determined as minima of the total energy; this necessarily leads to deviations of electronic wavefunctions and nuclei positions near the boundaries. In many cases, details of the nuclei positions and electronic states near the boundaries are complicated and not easily discussed on general grounds.
Thus, a general proof for $C_n$-symmetric insulators is desired. In the following, we give a general proof for this problem. 

In this paper, we assume that the charges of the respective ions are integer multiples of the electronic charge $|e|$. It naturally holds in real materials. On the other hand, some theoretical models may have non-integer ionic charges, but such models are outside of our theory.

\subsection{Derivation of the general corner charge formula}
We consider a crystal of finite size which preserves the $C_n$ symmetry. We assume that the system is insulating, in the sense that the Fermi level is within the energy gap for both the bulk and the boundaries. 
In some cases, degenerate corner states exist at the Fermi energy, which violates the above assumption, when charge neutrality is assumed. However, we can add or remove several electrons so that no states are at the Fermi energy. This deviation from charge neutrality is called filling anomaly\cite{PhysRevB.99.245151}.

\begin{figure}
  \centerline{\includegraphics[width=8cm,clip]{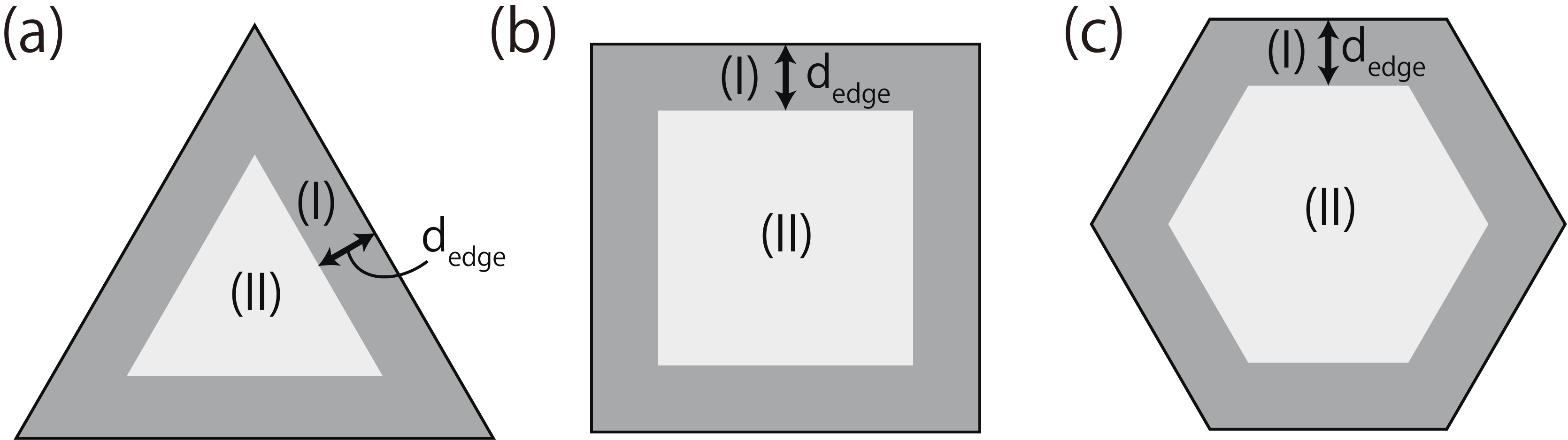}}
  \caption{(Color online) A conceptual picture of the region division for (a) $C_3$, (b) $C_4$ and (c) $C_6$-symmetric systems. In each system, the width of the region (I), $d_{\text{edge}}$, is taken to be large enough so that both the electronic states and the ionic positions in the region (II) are the same as those at the bulk. 
}
    \label{region_division}
\end{figure}

In order to calculate the filling anomaly, we divide the system conceptually into two $C_n$-symmetric regions (I) and (II), as shown in Fig.~{\ref{region_division}}. 
The regions (I) and (II) can be regarded as surface and bulk regions, respectively, and we explain this division in the following. The width of the surface region (I), $d_{\text{edge}}$, is taken to be large enough so that both the electronic states and the ionic positions in the region (II) are the same as those at the bulk. 

To calculate the filling anomaly, we proceed with the following steps.
First, we classify the ions into two groups according to which region they belong to, so that no ions are at the junctures of two regions.

Second, we classify the electronic states. We assume that the occupied states of this system can be described by a $C_n$-symmetric basis set $\{\phi_{i}\}$ composed of exponentially localized basis orbitals\cite{PhysRevB.48.4442}. Well within the crystal, the electronic states can be well described by bulk Wannier functions $W_{nl}$, and we take these Wannier functions $W_{nl}$ to be among $\{\phi_{i}\}$. The remaining $\phi_{i}$'s are additional localized basis orbitals $\varphi_i^{(\text{add})}$ near the boundary. We classify them into regions (I) and (II) so that the set of orbitals in region (I) and that in region (II) are $C_n$ symmetric. 
By construction the numbers of ions and electronic localized basis orbitals in the region (I) are integer multiples of $n$. 
Here, we impose a condition that all the additional localized basis orbitals $\varphi_i^{(\text{add})}$ are classified into the region (I), and we choose the region (I) large enough so that this condition is satisfied. Thus, even if the electronic states near the boundary are largely deformed, our theory is still applicable, by incorporating these deformed wavefunctions into region (I). We note that this construction of localized basis functions is similar to that used in the argument in Ref.~[\onlinecite{PhysRevB.48.4442}] to show that the bulk polarization in terms of the Zak phase gives surface charge density. 
We note that some degrees of freedom remain in this division into two regions, but they do not affect the following argument.


Third, we calculate the total charge of the finite crystal in terms of mod $n|e|$. First, the total charge contribution from the region (I) should be $0$ (mod $n|e|$) because the number of the nuclei and the electronic basis orbitals within region (I) is an integer multiple of $n$. 
Next, we calculate total charges of electrons and ions in the region (II). First, in the region (II), all the orbitals are Wannier orbitals $W_{nl}$, and therefore positions of the ions and electrons are classified in terms of Wyckoff positions. Because the whole crystal is $C_n$ symmetric, the center of the crystal should be located at the Wyckoff position with $C_n$ local symmetry. 
Let $1a$ be the Wyckoff position with $C_n$ symmetry located at the center of the crystal. Then, the ions not located at $1a$ contribute $nM|e|$ ($M\in\mathbb{Z}$) to the net charge in the region (II), while those at 1a contribute $nM^{\prime}|e|+n_{1a}^{(\text{ion})}|e|$ ($M^{\prime}\in\mathbb{Z}$), where $n_{1a}^{(\text{ion})}|e|$ is the charge of the ion at the $1a$ position. 
Therefore, the total ionic charge in the region (II) is equal to $n_{1a}^{(\text{ion})}|e|$ mod $n|e|$ and so is that in the whole crystal. Similarly, the total electron charge is equal to $-n_{1a}^{\text{(e)}}|e|$ (mod $n|e|$) where $n_{1a}^{\text{(e)}}$ is the number of electron Wannier orbitals at the $1a$ position. 
From the above discussion, the total charge of the system in terms of modulo $n|e|$ is calculated as follows, 
\begin{align}
Q_{\text{tot}}\equiv(n_{1a}^{\text{(ion)}}-n_{1a}^{\text{(e)}})|e| 
\quad (\text{mod}\ n|e|),
\label{Q_tot}
\end{align}
where $n_{1a}^{\text{(ion)}}|e|$ is an ionic charge at $1a$, and $n_{1a}^{\text{(e)}}$ is the number of electronic Wannier functions at $1a$. This formula represents a filling anomaly in the crystal.

\begin{figure}
  \centerline{\includegraphics[width=8cm,clip]{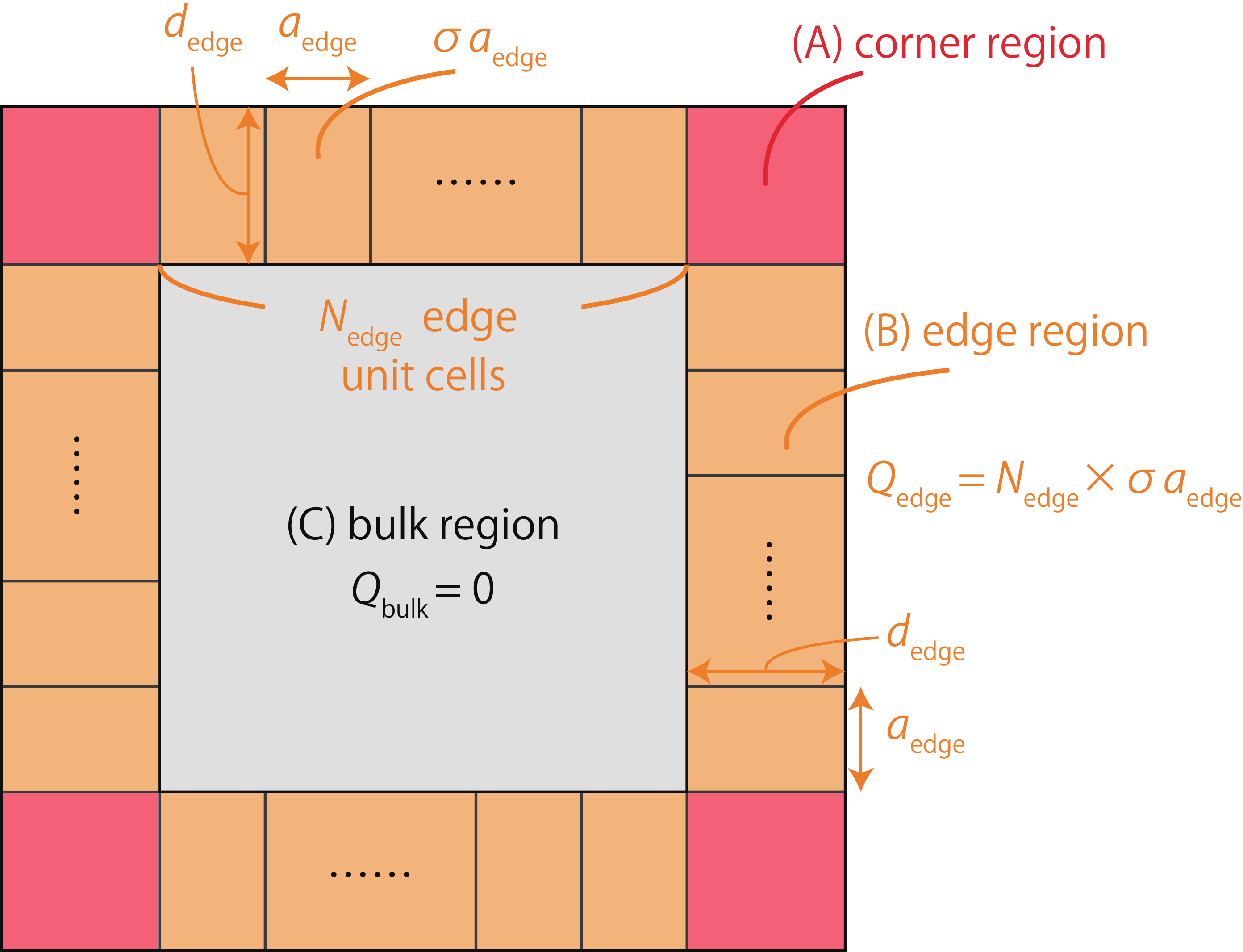}}
  \caption{(Color online) A conceptual picture of the region division for calculating the corner charge for a $C_4$-symmetric system as an example. The total charge in each region is calculated by integrating the moving average of the total charge density $\overline{\rho}(\bm{x})$ in each region. In the bulk region (C), $\overline{\rho}(\bm{x})=0$, and the total charge in the bulk region $Q_{\text{bulk}}$ is zero. 
The edge region (B) includes an integer number of edge unit cells. 
The charge of each edge unit cell is calculated by the product of the edge charge density $\sigma=\bm{P}\cdot\bm{n}$ and the edge period $a_{\text{edge}}$.
The total charge is represented as summation of the edge charges and the corner charges: $Q_{\text{tot}}=n(Q_{\text{edge}}+Q_{\text{corner}})$. Therefore, when $Q_{\text{edge}}$ vanishes, $Q_{\text{corner}}\equiv Q_{\text{tot}}/n$ (mod $e$). 
}
    \label{region_integral}
\end{figure}

Fourth, we discuss how $Q_{\text{tot}}$ is distributed in the crystal. As a first step, we define the moving average of the total charge density $\overline{\rho}(\bm{x})$ as follows:
\begin{align}
\overline{\rho}(\bm{x})=\frac{1}{|\Omega|}\int_{\Omega}\rho(\bm{x}+\bm{x}^{\prime})\ d\bm{x}^{\prime}. 
\end{align}
Here, $\rho(\bm{x})$ is the total charge density including ions and electrons, $\Omega$ is a $C_n$-symmetric unit cell whose center is at the origin and $|\Omega|$ is the area of $\Omega$. 
From the definition, $\overline{\rho}(\bm{x})$ have the following properties:
\begin{align}
\overline{\rho}(\bm{x})&=0 \quad (\bm{x}\in\text{bulk}), 
\label{rhobar1}
\\
\int_{\text{crystal}}d\bm{x}^{\prime}\ 
\overline{\rho}(\bm{x}^{\prime})\ &=Q_{\text{tot}}.
\label{rhobar2}
\\
\overline{\rho}(C_n\cdot\bm{x})&=\overline{\rho}(\bm{x})
\label{rhobar3}
\end{align}
We note that as long as $\overline{\rho}(\bm{x})$ satisfies Eqs.~(\ref{rhobar1}), (\ref{rhobar2}) and (\ref{rhobar3}), we can also use other definitions for moving averages. 
In a sufficiently large system, we expect that electronic states in the middle part of the edges far away from the corners are almost the same as those on an edge of a semi-infinite system with an open boundary condition in one direction. To calculate the corner charge based on this expectation, we divide the system into three types of regions: the (A) corner, (B) edge, and (C) bulk regions in a $C_n$-symmetric way as schematically shown in Fig.~\ref{region_integral}. The edge region (B) should be chosen to satisfy the following conditions. First, it should be sufficiently far away from the corners, such that the electronic states in the edge region (B) are the same as those on an edge of the corresponding semi-infinite system. 
Second, it should have a width equal to an integer multiple of the period along the edge, $N_{\text{edge}}a_{\text{edge}}$, and a sufficiently long depth into the bulk, $d_{\text{edge}}$, such that in the remaining bulk region (C), the electronic states are the same as those for a bulk infinite system. 
Here, we assume that one edge region includes $N_{\text{edge}}$ edge unit cells.
Since $\overline{\rho}(\bm{x})$ takes nonzero values only in the (A) corner and (B) edge regions, its integral can be calculated as the sum of the total charges in these regions:
\begin{align}
\int_{\text{crystal}}d\bm{x}^{\prime}\ 
\overline{\rho}(\bm{x}^{\prime})\ &=n\Bigl{(}Q_{\text{edge}}+Q_{\text{corner}}\Bigr{)},
\end{align}
where $Q_{\text{edge}}$ and $Q_{\text{corner}}$ are the total charges at one of the (B) edge regions and one of the (C) corner regions, respectively. 
Then $Q_{\text{edge}}$ can be calculated as:
\begin{align}
Q_{\text{edge}}
&=N_{\text{edge}}\sigma_{\text{edge}}a_{\text{edge}}.
\end{align}
Here, $N_{\text{edge}}$ is the number of edge unit cells within one edge region, $\sigma_{\text{edge}}$ is the edge charge density, and $a_{\text{edge}}$ is the period of the edge unit cell.

Finally, we calculate the corner charge. We assume that the edge is charge neutral, $\sigma_{\text{edge}}=0$. 
Then the extra charge $Q_{\text{tot}}$ is distributed into the $n$ corners. Since the crystal is assumed to be $C_n$-symmetric, the distributed charges are equal among the corners, and this gives a corner charge formula: 
\begin{align}
Q_{\text{corner}}=Q_{c\_1a}^{(n)}\equiv\frac{(n_{1a}^{\text{(ion)}}-n_{1a}^{\text{(e)}})|e|}{n} 
\quad (\text{mod}\ |e|). 
\label{Q_c_general}
\end{align}
Here, $Q_{c\_1a}^{(n)}$ means the corner charge when the center of the crystal is located at the $1a$ Wyckoff position.

If the edge charge density $\sigma_{\text{edge}}\neq0$, the extra charge $Q_{\text{tot}}$ is distributed not only to the corners, but also to the edges. 
Since the separation of charges between the corners and the edges depends on the choice of the corner regions, the corner charge is not well defined.

\subsection{Comparison with previous works}
Our discussion so far clarifies that the formula (\ref{Q_c_general}) for the corner charge applies to a general system, as long as the system is $C_n$-symmetric and is insulating over the whole system, irrespective of the system details.
In particular, our theory applies to (i) systems with boundaries with high Miller indices, (ii) those with surface reconstructions, and (iii) systems with nonzero $\bm{P}$ but with zero edge charge density $\sigma_{\text{edge}}=0$. In the following, we briefly explain these three cases. Details of (i) and (ii) are given in Sec.~IV and those of (iii) in Sec.~III.

First is the case (i) with boundaries with higher Miller indices. In the previous works, the arguments are limited to boundaries with low Miller indices such as (10) and (01) edges, but our theory applies to general Miller indices. For example, a $C_4$-symmetric square system with edges along (34) and ($4\bar{3}$) directions is covered by our theory.
Second, we can treat cases with (ii) boundaries with surface reconstructions. It is nontrivial because the surface reconstruction may make the perioidicity at edges different from that in the bulk. In some cases, as we discusss later, the edges without the surface reconstruction are gapless and $\sigma_{\text{edge}}$ is zero, but those with the surface reconstruction may become gapped, which makes our theory applicable.

Third, our theory covers also the cases (iii) with $\bm{P}\neq0$. Our assumption is $\sigma_{\text{edge}}=0$, which does not necessarily mean $\bm{P}=0$. 
In contrast, in the previous works\cite{PhysRevB.99.245151,PhysRevResearch.1.033074,PhysRevResearch.2.043131,PhysRevB.102.165120}, the bulk polarization $\bm{P}$ is assumed to be 0, which ensures that the edge charge density $\sigma_{\text{edge}}$ is 0. We find that this assumption of $\bm{P}=0$ in the previous works is too strong. The bulk polarization $\bm{P}$ is related to the edge charge density\cite{PhysRevB.48.4442,resta1992theory,PhysRevB.47.1651,RevModPhys.66.899} via 
\begin{align}
\sigma_{\text{edge}}\equiv\bm{P}\cdot\bm{n}\ (\text{mod}\ |e|/a_{\text{edge}}),
\end{align}
where $\bm{n}$ is the unit normal vector of the edge. It is not an equality, because the bulk polarization $\bm{P}$ is defined only in terms of modulo some unit \cite{PhysRevB.48.4442, resta1992theory,PhysRevB.47.1651,RevModPhys.66.899}, which we call polarization quantum. Thus, our assumption $\sigma_{\text{edge}}=0$ means $\bm{P}\cdot\bm{n}\equiv0$ (mod $|e|/a_{\text{edge}}$), which does not
necessarily mean $\bm{P}=0$.

\subsection{Corner-charge formula in terms of rotation eigenvalues}
In this section, we rewrite our formula for the corner charge in terms of rotation eigenvalues of the Bloch eigenstates at high-symmetry wavevectors following Refs.~[\onlinecite{PhysRevB.99.245151},\onlinecite{PhysRevResearch.1.033074},\onlinecite{PhysRevResearch.2.043131}], in three different systems 
with and without time-reversal symmetry, and with and without spin-orbit coupling, separately. We need to rewrite it, because eigenstates in a crystal are Bloch wavefunctions, extended over the crystal, and one cannot directly calculate $n_{1a}^{(e)}$ from the Bloch wavefunctions. Then, we discuss a relation between our formula and the formulas in the previous works\cite{PhysRevB.99.245151,PhysRevResearch.1.033074,PhysRevResearch.2.043131}. 
The notable difference from the previous works is that the Wyckoff positions of the ions are not always 1a, and the corner charge generally depends on the filling $\nu$ as in Ref.~[\onlinecite{PhysRevResearch.2.043131}]. The detail of the calculation is shown in the Appendix.

By following Refs.~[\onlinecite{PhysRevB.99.245151},\onlinecite{PhysRevResearch.1.033074},\onlinecite{PhysRevResearch.2.043131}], 
we introduce topological invariants which distinguish the Wannier functions with different Wyckoff positions. 
With a sufficient number of topological invariants, there can be a one-to-one correspondence between the values of the topological invariants and Wyckoff positions of the occupied Wannier orbitals. We assume that the occupied bands can be written in terms of Wannier functions. Then, the number of Wannier orbitals at each Wyckoff position is determined from the topological invariant for the occupied bands.

\begin{figure}[t]
  \centerline{\includegraphics[width=8cm,clip]{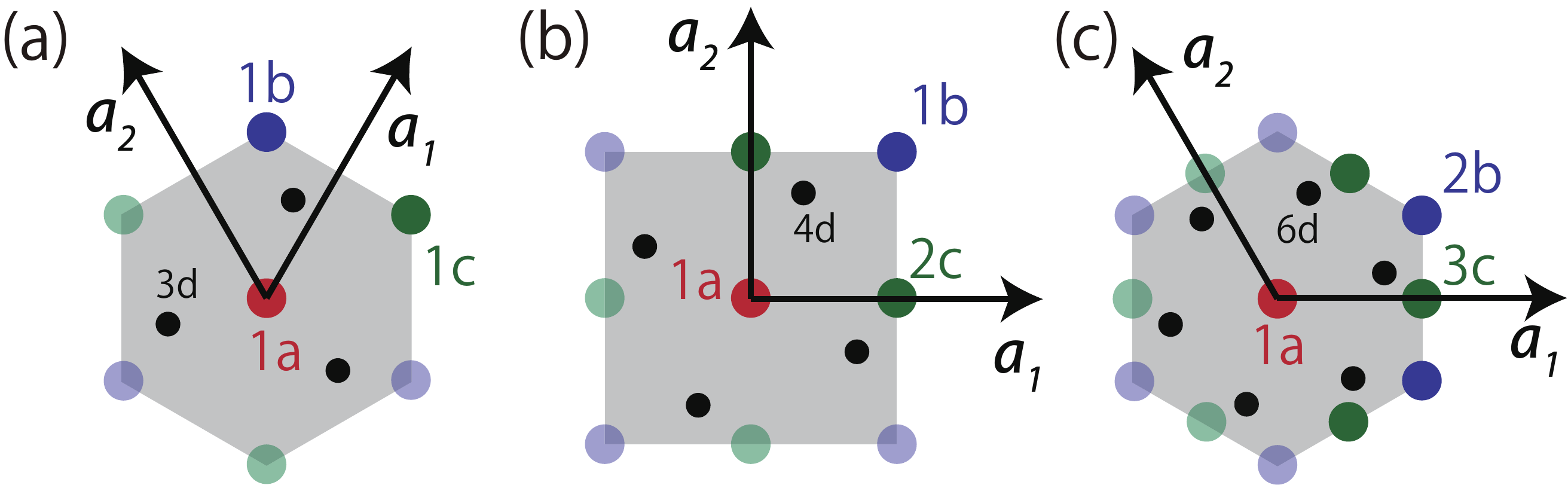}}
  \caption{(Color online)  Wyckoff positions in (a) $C_3$, (b) $C_4$, (c) $C_6$-symmetric systems. The points with the same colors belong to the same Wyckoff positions. Here, $\bm{a}_1$ and $\bm{a}_2$ are primitive translation vectors, and (a) $\bm{a}_1=(\frac{a}{2},\frac{\sqrt{3}}{2}a)$, $\bm{a}_2=(-\frac{a}{2},\frac{\sqrt{3}}{2}a)$, (b) $\bm{a}_1=(a,0)$, $\bm{a}_2=(0,a)$, (c) $\bm{a}_1=(a,0)$, $\bm{a}_2=(-\frac{a}{2},\frac{\sqrt{3}}{2}a)$. 
}
    \label{Cn_Wyckoff}
\end{figure}

Figure {\ref{Cn_Wyckoff}} shows the Wyckoff positions under $C_n$ rotational symmetry ($n=3,4,6$). In each system with $C_n$ symmetry, there are four kinds of Wyckoff positions. We define the number of occupied Wannier functions at Wyckoff position $mX$ as $n_{mX}^{(e)}$. Here, $X=a,b,c,d$ represent the types of the Wyckoff positions, and $m$ represents the multiplicity of a Wyckoff position. The weighted sum of the $n_{mX}^{(e)}$ is equal to the electron filling, i.e. the number of occupied bands $\nu$:
\begin{align}
\sum_{X} m n_{mX}^{(e)}=\nu. 
\end{align}
For the calculation of the corner charge, the value of $n_{1a}^{(e)}$ is needed. As we show in the appendix, $mn_{mX}^{(e)}$ modulo $n$ ($X\neq a$) is directly calculated from the rotational topological invariants defined from rotation eigenvalues of the Bloch wavefunctions, and $n_{1a}^{(e)}$ is determined from $\nu-\sum_{X\neq a} m n_{mX}^{(e)}$. In the appendix, we show the detailed calculation of $n_{mX}^{(e)}$.

\subsubsection{class A}
First, we consider a system called class A. 
Unlike the class AI systems\cite{PhysRevB.99.245151,PhysRevResearch.2.043131} or class AII systems\cite{PhysRevResearch.1.033074,kooi2021bulk} considered in previous studies, the class A systems do not have time-reversal symmetry. 
From Eq.~(\ref{Q_c_general}) and Appendix A, the corner charge is calculated as follows:
\begin{align}
Q_{c\_1a}^{(3)}
&\equiv\frac{|e|}{3}\Big{(}n_{1a}^{(\text{ion})}-\nu-[K_1^{(3)}]-[K_2^{(3)}]-[K_1^{\prime(3)}]-[K_2^{\prime(3)}]\Big{)} 
\notag \\
&\ \quad\quad\quad\quad\quad\quad\quad\quad\quad\quad\quad\quad\quad\quad\quad\ (\text{mod}\ e),
\label{Qc_3_A}
\\
Q_{c\_1a}^{(4)}
&\equiv\frac{|e|}{4}\Big{(}n_{1a}^{(\text{ion})}-\nu+[X_1^{(2)}]-\frac{1}{2}[M_1^{(4)}]+\frac{3}{2}[M_3^{(4)}]\Big{)}
\notag \\
&\ \quad\quad\quad\quad\quad\quad\quad\quad\quad\quad\quad\quad\quad\quad\quad\ (\text{mod}\ e),
\label{Qc_4_A}
\\
Q_{c\_1a}^{(6)}
&\equiv\frac{|e|}{6}\Big{(}n_{1a}^{(\text{ion})}-\nu+2[K_1^{(3)}]+\frac{3}{2}[M_1^{(2)}]
\Big{)}\ (\text{mod}\ e).
\label{Qc_6_A}
\end{align}
Here, the rotation topological invariant $[\Pi_{p}^{(n)}]$ represents the difference of the number of states with $C_n$ eigenvalue $e^{\frac{2\pi(p-1)i}{n}}$ for spinless systems, and $e^{\frac{2\pi(p-1)i}{n}}e^{\frac{\pi i}{n}}$ for spinful systems, between the rotation-invariant $\bm{k}$-points $\Pi$ and $\Gamma$. We assumed $(C_n)^n=1$ and $(C_n)^n=-1$ for spinless and spinful systems, respectively.

In addition to Eqs.~(\ref{Qc_3_A})-(\ref{Qc_6_A}), where the center of the $C_n$-symmetric crystal is at $1a$, in appendix A, we also discuss the formula when the center of the crystal is at other Wyckoff positions, i.e., $1b$ and $1c$ for $C_3$ and $1b$ for $C_4$. The formulae are summarized as follows:
\begin{align}
Q_{c\_1b}^{(3)}
&\equiv\frac{|e|}{3}\Big{(}n_{1b}^{(\text{ion})}+[K_1^{(3)}]+[K_2^{\prime(3)}]\Big{)}\ (\text{mod}\ e),
\\
Q_{c\_1c}^{(3)}
&\equiv\frac{|e|}{3}\Big{(}n_{1c}^{(\text{ion})}+[K_2^{(3)}]+[K_1^{\prime(3)}]\Big{)}\ (\text{mod}\ e),
\\
Q_{c\_1b}^{(4)}
&\equiv\frac{|e|}{4}\Big{(}n_{1b}^{(\text{ion})}-[X_1^{(2)}]+\frac{3}{2}[M_1^{(4)}]-\frac{1}{2}[M_3^{(4)}]
\Big{)}
\notag \\
&\ \quad\quad\quad\quad\quad\quad\quad\quad\quad\quad\quad\quad\quad\quad\ (\text{mod}\ e).
\end{align}

\subsubsection{class AI}
Here, we consider a spinless system with time-reversal symmetry called class AI. In this case, there are some constraints on the topological invariants $[\Pi_{p}^{(n)}]$, and the corner charge formula is somewhat simplified. 

\begin{align}
Q_{c\_1a}^{(3)}
&\equiv\frac{|e|}{3}(n_{1a}^{(\text{ion})}-\nu)-\frac{|e|}{3}[K_1^{(3)}]\ (\text{mod}\ e),
\label{Qc3_AI_1a}
\\
Q_{c\_1a}^{(4)}
&\equiv\frac{|e|}{4}\Big{(}n_{1a}^{(\text{ion})}-\nu+[X_1^{(2)}]-2[M_1^{(4)}]-3[M_2^{(4)}]
\Big{)} 
\notag \\
&\ \quad\quad\quad\quad\quad\quad\quad\quad\quad\quad\quad\quad\quad\quad\quad\ (\text{mod}\ e),
\label{Qc4_AI_1a}
\\
Q_{c\_1a}^{(6)}
&\equiv\frac{|e|}{6}\Big{(}n_{1a}^{(\text{ion})}-\nu+2[K_1^{(3)}]+\frac{3}{2}[M_1^{(2)}]
\Big{)}\ (\text{mod}\ e).
\label{Qc6_AI_1a}
\end{align}

In particular, these formulae are reproduced to the results in Ref.~[\onlinecite{PhysRevB.99.245151}] only if all the ions are located at the center of the $C_n$-symmetric unit cell (i.e., Wyckoff position $1a$) leading to $n_{1a}^{(\text{ion})}=\nu$ and if the electric polarization is zero.  
The formulae are reduced to 
Eq.~(11) of Ref.~[\onlinecite{PhysRevB.99.245151}], i.e.,
\begin{align}
Q_{c\_1a}^{(3)}
&\equiv-\frac{|e|}{3}[K_2^{(3)}]
\ (\text{mod}\ e),
\label{Qc3_AI_1a_Benalcazar}
\\
Q_{c\_1a}^{(4)}
&\equiv-\frac{|e|}{4}
\Big{(}[X_1^{(2)}]+2[M_1^{(4)}]+3[M_2^{(4)}]\Big{)}\ (\text{mod}\ e),
\label{Qc4_AI_1a_Benalcazar}
\\
Q_{c\_1a}^{(6)}
&\equiv-\frac{|e|}{4}[M_1^{(2)}]-\frac{|e|}{6}[K_1^{(3)}]\ (\text{mod}\ e).
\label{Qc6_AI_1a_Benalcazar}
\end{align}

As shown in appendix B, we also discuss the formula when the center of the $C_n$-symmetric crystal is not at the Wyckoff position $1a$. The formulae are summarized as follows:
\begin{align}
Q_{c\_1b}^{(3)}
&\equiv\frac{|e|}{3}\Big{(}n_{1b}^{(\text{ion})}-[K^{\prime(3)}_2]\Big{)}\ (\text{mod}\ e),
\\
Q_{c\_1c}^{(3)}
&\equiv\frac{|e|}{3}\Big{(}n_{1c}^{(\text{ion})}-[K^{(3)}_2]\Big{)}\ (\text{mod}\ e),
\\
Q_{c\_1b}^{(4)}
&\equiv\frac{|e|}{4}\Big{(}n_{1b}^{(\text{ion})}-[X^{(2)}_1]+2[M^{(4)}_1]+[M^{(4)}_2]
\Big{)}
\notag \\
&\ \quad\quad\quad\quad\quad\quad\quad\quad\quad\quad\quad\quad\quad\quad\ (\text{mod}\ e).
\end{align}

\subsubsection{class AII}
Here, we consider a spinful system with time-reversal symmetry called class AII. In Refs.~[\onlinecite{PhysRevResearch.1.033074}, \onlinecite{kooi2021bulk}], in the class AII systems, the corner charge is defined modulo $2e$ due to the Kramers theorem. However, as pointed out in Ref.~[\onlinecite{PhysRevB.102.165120}], we argue that it should be defined modulo $e$ in general cases. For example, the nuclei can have an odd number of charges in the unit of electronic charge $|e|$, which leads to a change of the corner charge by $e$ without changing the bulk. 
It occurs when some surface adatoms are allowed, as discussed in Ref.~[\onlinecite{PhysRevB.102.165120}]. 
It is also the case in compounds containing elements with odd atomic numbers; a change in crystal terminations results in a change of the corner charge by $e$.  
For the above reasons, we deal with corner charges defined by modulo $e$, even in the class AII.

As with class AI systems, the topological invariants $[\Pi_{p}^{(n)}]$ have some constraints due to time-reversal symmetry, and the corner charge formula is somewhat simplified. 
\begin{align}
Q_{c\_1a}^{(3)}
&=\frac{|e|}{3}(n_{1a}^{(\text{ion})}-\nu)-\frac{|e|}{3}[K_2^{(3)}]
\ (\text{mod}\ e),
\\
Q_{c\_1a}^{(4)}
&\equiv\frac{|e|}{4}(n_{1a}^{(\text{ion})}-\nu)+\frac{|e|}{2}[M_1^{(4)}]
\ (\text{mod}\ e), 
\\
Q_{c\_1a}^{(6)}
&=\frac{|e|}{6}(n_{1a}^{(\text{ion})}-\nu)+\frac{|e|}{3}[K_1^{(3)}]\ (\text{mod}\ e).
\end{align}

In appendix C, we also discuss the formula when the center of the $C_n$-symmetric crystal is not at the Wyckoff position $1a$. The formulae are summarized as follows:
\begin{align}
Q_{c\_1b}^{(3)}
&\equiv\frac{|e|}{3}\Big{(}n_{1b}^{(\text{ion})}-[K^{(3)}_1]\Big{)}\ (\text{mod}\ e),
\\
Q_{c\_1c}^{(3)}
&\equiv\frac{|e|}{3}\Big{(}n_{1c}^{(\text{ion})}-[K^{\prime(3)}_1]\Big{)}\ (\text{mod}\ e),
\\
Q_{c\_1b}^{(4)}
&\equiv\frac{|e|}{4}\Big{(}n_{1b}^{(\text{ion})}-2[M^{(4)}_1]\Big{)}\ (\text{mod}\ e).
\end{align}

\section{Corner charge with nonzero bulk polarization}

In this section, we consider cases with nonzero bulk polarization $\bm{P}$. In the previous works\cite{
PhysRevB.99.245151,
PhysRevResearch.1.033074,
PhysRevResearch.2.043131,
PhysRevB.102.165120
}, the bulk polarization $\bm{P}$ is set to be 0, to ensure that the edge charge density per edge unit cell is 0 modulo $|e|$. However, as we explain in the following, even if the bulk polarization $\bm{P}$ is nonzero, as long as the edge charge density $\sigma_{\text{edge}}$ is 0, our general formula is applicable. 

To illustrate this, we consider a $C_4$-symmetric tight-binding model with nonzero bulk polarization as an example. We show that the fractional corner charge appears even though the bulk polarization is nonzero, as long as the edge charge density is 0. Moreover, we also show that two finite $C_4$-symmetric flakes with different terminations can have different corner charges due to difference in the $C_4$ centers. 
This property is specific to cases with nonzero polarization.

\begin{figure}
  \centerline{\includegraphics[width=8cm,clip]{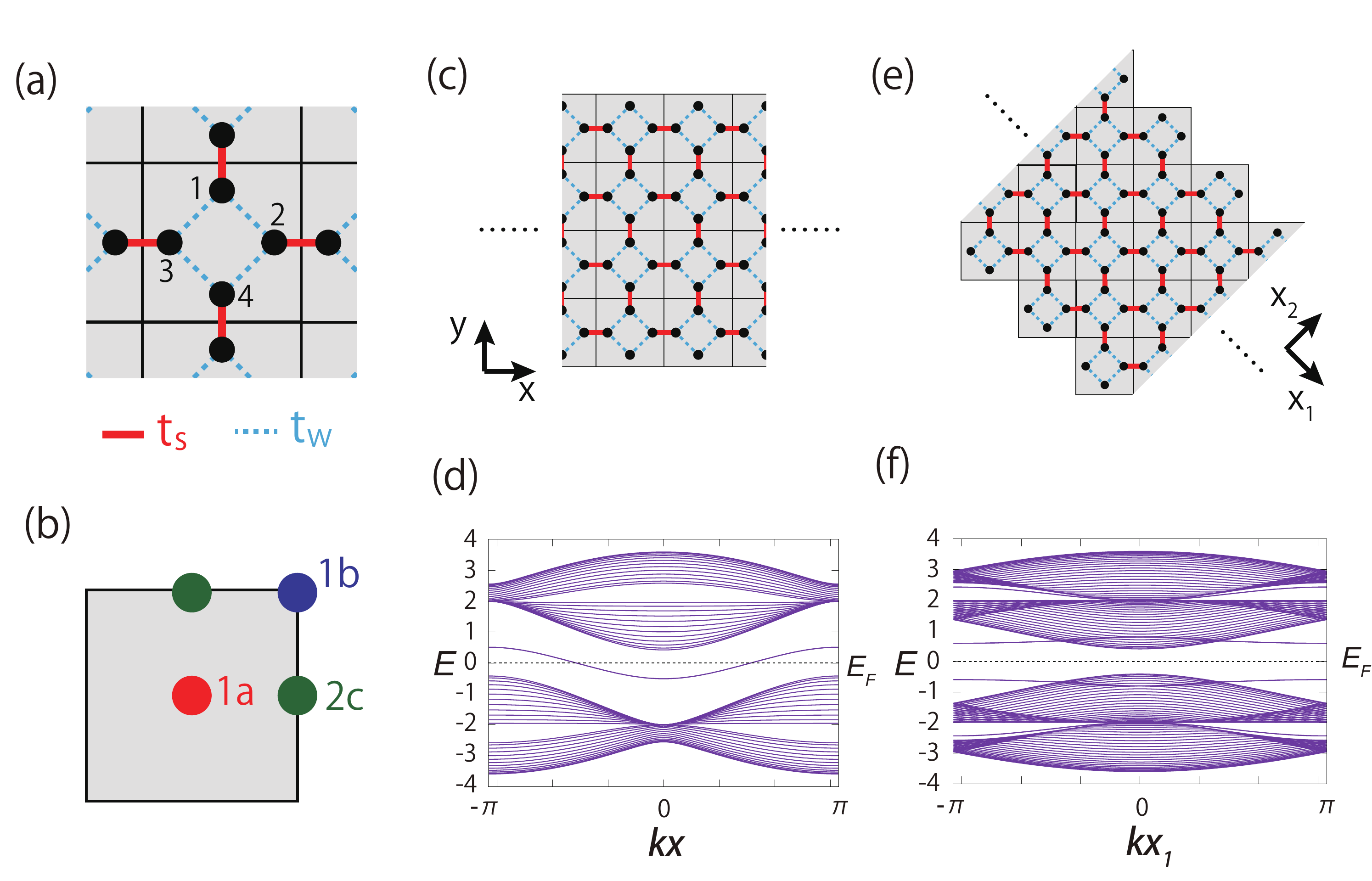}}
  \caption{(Color online) Numerical calculation for the $C_4$-symmetric tight-binding model in the ribbon geometry. (a) Hopping terms of the model. The hopping amplitude is $t_s$ ($t_w$) for bonds colored in red (blue). We set $t_s=2$ and $t_w=0.8$ in the numerical calculation. (b) Wyckoff positions of the $C_4$ symmetric system. (c,d) The ribbon system cut along (01) direction and its energy spectrum. Due to the nonzero electric polarization, fractional edge charge appears, and the edge energy spectrum is metallic. (e,f) The ribbon system cut along (11) direction and its energy spectrum. Unlike that along (01) direction, the edge energy spectrum is gapped. }    
\label{C4_numerical}
\end{figure}

Our model is a simple spinless tight-binding model. It has background positive ions with $2|e|$ charge at $1a$, being the center of the square unit cell (see Fig.~{\ref{C4_numerical}}(a),(b)). The Hamiltonian is given as $\hat{H}=\sum_{\langle ij\rangle}t_{ij}c_i^{\dagger}c_j$, where $t_{ij}=t_s\ (t_w)$ for inter-unit-cell (intra-unit-cell) bonds colored in red (blue) in Fig.~{\ref{C4_numerical}}(a). We set $t_s=2$ and $t_w=0.8$ in the numerical calculation. The bulk bands are gapped, and we set the Fermi energy $E_{F}=0$ within the gap. The parameters are chosen so that the occupied Wannier orbitals are located at $2c$ Wyckoff positions, in the middle of the red bonds. In this example, the bulk polarization is $\bm{P}=(e/2a)(1,1)$ (mod $e/a$), where $a$ is the lattice constant. Therefore, if we cut the system along the $x$ and $y$ axis as shown in Fig.~{\ref{C4_numerical}}(c), nonzero fractional edge charge appears, $\sigma_{\text{edge}} a_{\text{edge}}=\bm{P}\cdot(1,0)a=\frac{e}{2}$, where the size of the edge unit cell $a_{\text{edge}}$ is taken to be equal to $a$. This model has half-filled edge states within the bulk gap, and they are metallic (see Fig.~{\ref{C4_numerical}}(d)). Thus, this is out of the scope of the corner charge formula. On the other hand, if the system is cut along the $(11)$ direction as shown in Fig.~{\ref{C4_numerical}}(e), the edge charge is 0 modulo $e$: $\sigma_{\text{edge}} a_{\text{edge}}=\bm{P}\cdot\bm{n}\sqrt{2}a\equiv0$ (mod $e$) per edge unit cell, where $\bm{n}=(1/\sqrt{2},1/\sqrt{2})$ and $a_{\text{edge}}=\sqrt{2}a$. Then the edge is insulating as shown in Fig.~{\ref{C4_numerical}}(f), and we can apply our general formula. 


In $C_4$-symmetric systems, there are two $C_4$-symmetric Wyckoff positions, $1a$ and $1b$. As shown in Fig.~{\ref{Cn_Wyckoff}}(b), their positions are $\bm{r}_{1a}=\bm{0}$ and $\bm{r}_{1b}=\frac{1}{2}\bm{a}_1+\frac{1}{2}\bm{a}_2$, respectively. Therefore, which formula to be used depends on whether the $C_4$ center of the finite crystal is $1a$ or $1b$. If the center of the finite crystal is at $1b$, we need to use a formula with $1a$ replaced by $1b$ in Eq.~(\ref{Q_c_general}): 
\begin{align}
Q_{c\_1b}^{(4)}
\equiv\frac{(n_{1b}^{\text{(ion)}}-n_{1b}^{\text{(e)}})|e|}{4} 
\quad (\text{mod}\ |e|). 
\label{Q_c_general_2}
\end{align}
The difference between Eqs.~(\ref{Q_c_general}) and (\ref{Q_c_general_2})
is expressed as 
\begin{align}
Q_{c\_1a}^{(4)}-Q_{c\_1b}^{(4)}
&=\frac{q_a-q_b}{4}
=\frac{-2q_b-2q_c-4q_d}{4}
\notag \\
&\equiv\frac{q_b+q_c}{2}
\quad (\text{mod}\ e).
\label{DeltaQ}
\end{align}
Here, $q_X=(n_{mX}^{\text{(ion)}}-n_{mX}^{\text{(e)}})|e|$ ($mX=1a,1b,2c,4d$). 
The r.h.s of Eq.~(\ref{DeltaQ}) is equal to 
$\bm{P}\cdot(a,0)$ and $\bm{P}\cdot(0,a)$. Therefore, when the quantized polarization along (10) or (01) direction is nonzero, the corner charge is different between a $C_4$-symmetric crystal with its center at $1a$ and one at $1b$.

We note that in Ref.~[\onlinecite{PhysRevB.102.165120}], systems without polarization are studied, and 
\begin{align}
\frac{1}{4}q_a&=\frac{1}{4}q_b\quad (\text{mod}\ e)\quad (n=4),\label{Watanabe_C4}
\end{align}
is shown, meaning that the corner charge is independent of whether the crystal center is at $1a$ or $1b$. 
It is a special case of our result.

\begin{figure}
  \centerline{\includegraphics[width=8cm,clip]{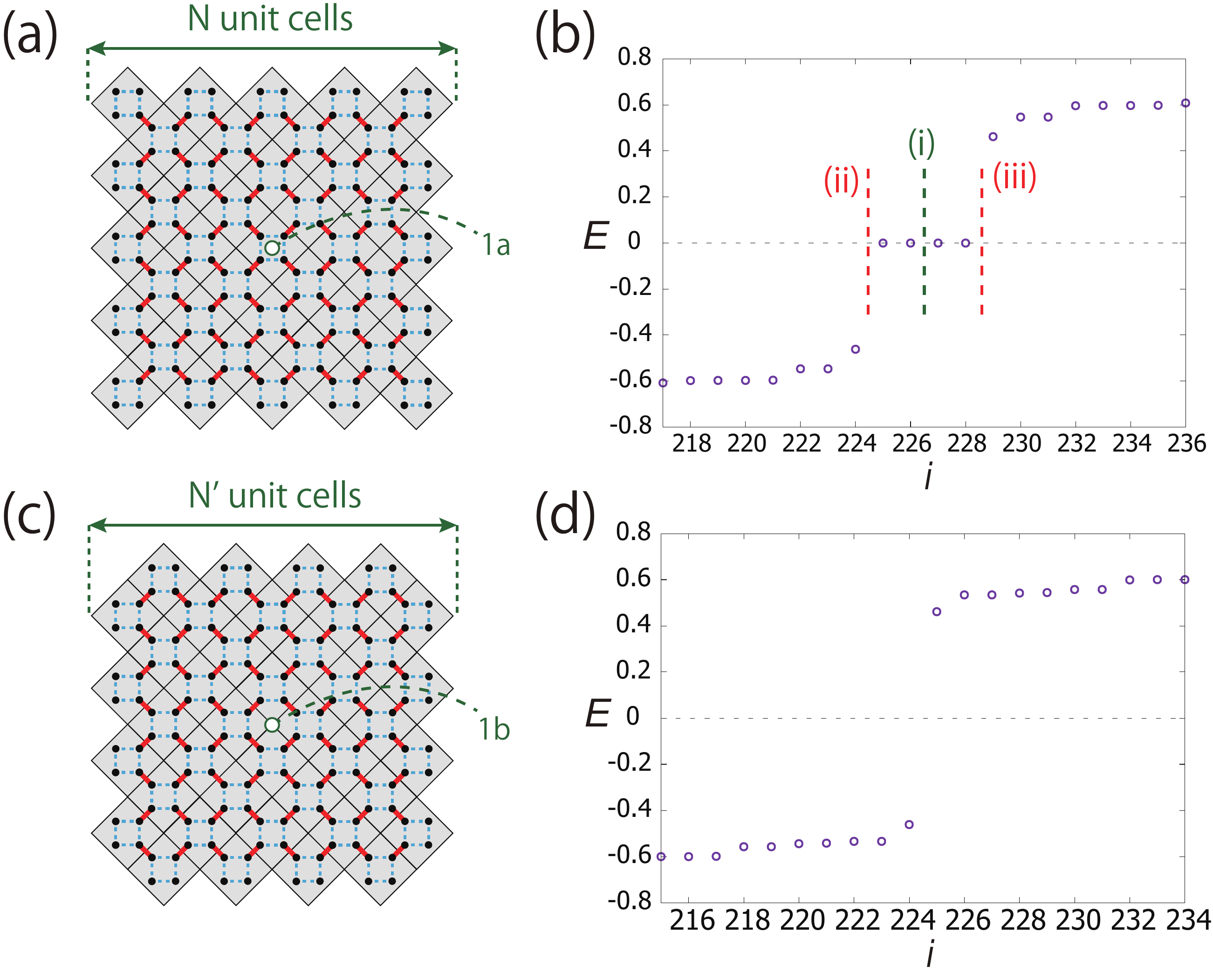}}
  \caption{(Color online) Numerical calculation for the $C_4$-symmetric tight-binding model in the flake geometry. (a,b) The finite system with the $C_4$ center at $1a$ and its energy spectrum. In the numerical calculation, the system size is $N=8$, and it contains $113$ unit cells in total. 
The charge-neutral filling shown by line (i) in the figure makes the system metallic. In an insulating filling as shown by lines (ii) and (iii), the total charge deviates from charge neutrality by $\pm2|e|$. As discussed in the main text, the nonzero total charge is equally distributed to the four corners, resulting in a corner charge $|e|/2$ (mod $|e|$). (c,d)  The finite system with the $C_4$ center at $1b$ and its energy spectrum. In the numerical calculation, the system size is $N^{\prime}=8$, and it contains $112$ unit cells in total.} 
\label{C4_numerical_corner}
\end{figure}

This difference between $Q_{c\_1a}^{(4)}$ and $Q_{c\_1b}^{(4)}$ can be attributed to a difference in the shape of the corners depending on the $C_4$ center. For example, when the shape of the crystal is as shown in Fig.~{\ref{C4_numerical_corner}}(a), the $C_4$ center is at $1a$, and the corner charge is calculated as $Q_{c\_1a}^{(4)}=(n_{1a}^{\text{(ion)}}-n_{1a}^{(e)})|e|/4=|e|/2$. On the other hand, when the crystal shape is as shown in Fig.~{\ref{C4_numerical_corner}}(c), the $C_4$ center is at $1b$, and the corner charge is calculated as $Q_{c\_1b}^{(4)}=(n_{1b}^{\text{(ion)}}-n_{1b}^{(e)})|e|/4=0$.

Correspondingly, in the system shown in Fig.~{\ref{C4_numerical_corner}}(a), where the $C_4$ center is at $1a$, four in-gap corner states appear (see Fig.~{\ref{C4_numerical_corner}}(b)). Here, we briefly explain the relation of the in-gap corner states and the corner charge. 
In the case of a charge-neutral filling indicated by the line (i) in Fig.~{\ref{C4_numerical_corner}}(b), the corner states are at the Fermi energy. In this case, we cannot define its corner charge. By adding or extracting two electrons as shown by lines (ii) and (iii), the whole system becomes gapped and the total charge deviates from charge neutrality by $\pm2|e|$. As discussed in Sec.~I\hspace{-.1em}I, the nonzero total charge is equally distributed to the four corners, resulting in a corner charge $|e|/2$ (mod $|e|$).
In contrast, in the system in Fig.~{\ref{C4_numerical_corner}}(c), where the $C_4$ center is at $1b$, there are no corner states (see Fig.~{\ref{C4_numerical_corner}}(d)). This agrees with our result in Eq.~(\ref{DeltaQ}).

In $C_3$-symmetric systems, there are three $C_3$-symmetric Wyckoff positions, $1a$, $1b$ and $1c$.  
As shown in Fig.~{\ref{Cn_Wyckoff}}(a), their positions are $\bm{r}_{1a}=\bm{0}$, $\bm{r}_{1b}=\frac{1}{3}\bm{a}_1+\frac{1}{3}\bm{a}_2$ and $\bm{r}_{1c}=\frac{2}{3}\bm{a}_1+\frac{2}{3}\bm{a}_2$, respectively. Therefore, similar to the case of $C_4$-symmetric systems, the formula for the corner charge depends on whether the $C_3$ center of the finite crystal is $1a$, $1b$ or $1c$, and let $Q^{(3)}_{c\_1a}$, $Q^{(3)}_{c\_1b}$, $Q^{(3)}_{c\_1c}$ represent the corresponding corner charges  respectively. Then, we obtain
\begin{align}
Q^{(3)}_{c\_1a}-Q^{(3)}_{c\_1b}
&=\frac{q_a-q_b}{3}
=\frac{-2q_b-q_c-3q_d}{3}
\notag \\
&\equiv\frac{q_b+2q_c}{3}\ (\text{mod}\ e),
\label{DeltaQ_3_1}
\\
Q^{(3)}_{c\_1a}-Q^{(3)}_{c\_1c}
&=\frac{q_a-q_c}{3}
=\frac{-q_b-2q_c-3q_d}{3}
\notag \\
&\equiv-\frac{q_b+2q_c}{3}\ (\text{mod}\ e). 
\label{DeltaQ_3_2}
\end{align}
As discussed in Ref.~[\onlinecite{PhysRevB.102.165120}], the r.h.s of Eq.~(\ref{DeltaQ_3_1}) is equal to both the electric polarization $|\Omega|\bm{P}\cdot\frac{\bm{b}_1}{2\pi}$ and $|\Omega|\bm{P}\cdot\frac{\bm{b}_2}{2\pi}$, where $\bm{b}_i$ ($i=1,2$) is the reciprocal lattice vector defined as $\bm{a}_i\cdot\bm{b}_j=2\pi\delta_{ij}$. Therefore, when the quantized polarization along $\bm{b}_1$ or $\bm{b}_2$ is nonzero, the corner charge depends on the $C_3$-center of the finite crystal. In particular, when the polarization is zero, $\frac{q_a}{3}=\frac{q_b}{3}=\frac{q_c}{3}$ (mod $e$) holds as shown in Ref.~[\onlinecite{PhysRevB.102.165120}], and the corner charge is independent of the $C_3$-center.

In a $C_6$-symmetric system, there is only one $C_6$-symmetric Wyckoff position, $1a$. In this case, the corner charge formula is unique.

We discuss the role of a choice of the unit cell in our theory. In the definition of the bulk polarization according to the modern theory of polarization \cite{PhysRevB.48.4442}, a choice of the unit cell plays a crucial role, and so does it in the present theory of the corner charge. It is noted that 
a choice of the unit cell is a gauge degree of freedom, which does not affect the system itself. The edge charge density $\sigma_{\text{edge}}$ and the corner charge $Q_c$ are observables, and therefore they cannot depend on a choice of the unit cell.

We stress that in our theory, the unit cell is always the primitive one in the bulk. Meanwhile, one can optionally take a unit cell larger than the primitive one. It has been discussed previously in the context of reducing the cases with nontrivial polarization into those with $\bm{P}\equiv 0$, in order to apply the previous theories based on $\bm{P}=0$ \cite{PhysRevResearch.1.033074,PhysRevB.102.165120}. Nonetheless, we note that taking a larger unit cell than the primitive one is misleading in some cases. For example, in the $C_4$-symmetric model in Fig.~\ref{C4_numerical}(a), the polarization $\bm{P}$ is equal to $\bm{P}=\frac{e}{2a}(1,1)$ (mod $e/a$). Meanwhile, if we take a larger unit cell of $2a\times2a$ instead of the primitive one, $a\times a$, $\bm{P}$ is treated in terms of modulo $e/(2a)$, and $\bm{P}\equiv 0$ (mod $e/(2a)$). Thus, with an enlarged unit cell, one cannot distinguish between the two possible cases for $C_4$-symmetric cases, i.e. $\bm{P}=\frac{e}{2a}(1,1)$ (mod $e/a$) and $\bm{P}=0$ (mod $e/a$). In terms of the edge charge density, the polarization quantum corresponds to the charge $|e|$ per one period along the edge. Thus in the present model, when $\bm{n}=(1,0)$, the edge charge per one edge periodicity is $\sigma_{\text{edge}} a_{\text{edge}}\equiv \bm{P}\cdot\bm{n} a_{\text{edge}}=\frac{e}{2}$ (mod $e$) when the edge periodicity taken to be $a_{\text{edge}}=a$. Meanwhile by enlarging the unit cell to $2a\times 2a$, the edge charge density becomes $\sigma_{\text{edge}} a_{\text{edge}}^{\prime}=e $, where $a_{\text{edge}}^{\prime}=2a$, and it is congruent to zero modulo $e$, corresponding to the trivial bulk polarization.
Here we note that the edge charge density $\sigma_{\text{edge}}$ is a physical observable, whereas the bulk polarization $\bm{P}$ is not an observable in a strict sense, due to its ambiguity modulo 
a polarization quantum. Thus it is not a contradiction that the bulk polarization $\bm{P}$ changes from nontrivial to trivial only by a change of the unit cell, being a gauge transformation.

Our theory, where the unit cell is always taken to be the primitive cell, is simple and convenient, because we do not need to switch the choice of the unit cell in the middle of the calculation. It is natural to work always on the primitive unit cell, because the choice of the unit cell, being a gauge degree of freedom, should not affect physical observables. One needs to be careful in dealing with the bulk polarization $\bm{P}$, because it is not an observable in a strict sense as we mentioned earlier. There might be a belief that the corner-charge quantization requires $\bm{P}=0$, as has been adopted in previous papers, but our theory has revealed that the correct condition necessary for the corner-charge quantization is $\sigma_{\text{edge}}=0$ and not $\bm{P}=0$.

\section{General edge orientations and surface reconstructions}

In this section, we consider edge charge densities for edges with general orientations. When the edge charge density is nonzero, the corner charge is nonzero. Nonetheless, even in these cases, surface reconstructions may make the edge density to vanish, and the corner charge becomes quantized. In this section, we discuss this in general systems.

\subsection{surface reconstruction}
\begin{figure}
  \centerline{\includegraphics[width=8cm,clip]{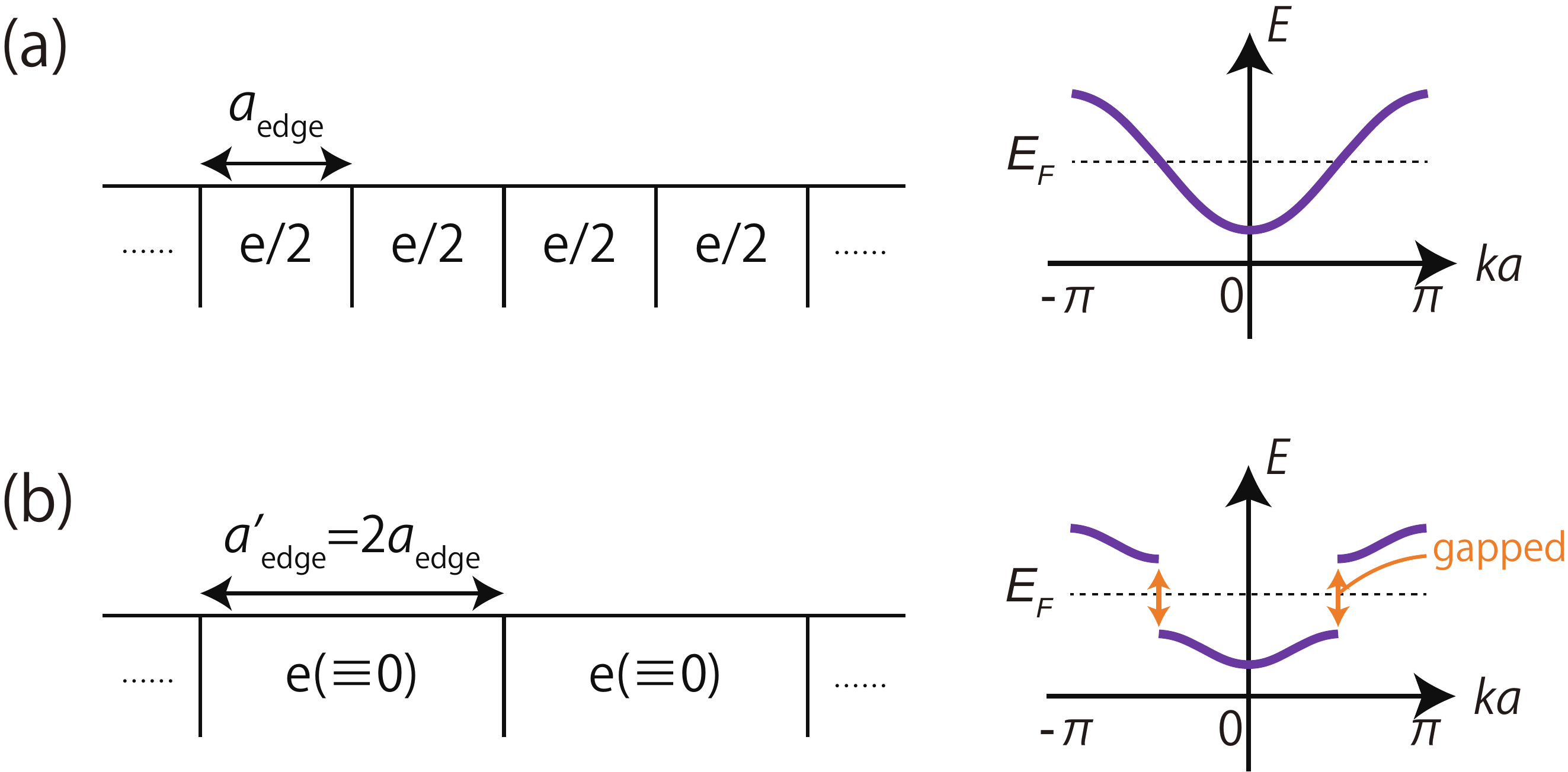}}
  \caption{(Color online) Conceptual picture of the surface reconstruction and corresponding edge band structures. (a) Before the surface reconstruction, each edge unit cell contains $\sigma_{\text{edge}}a_{\text{edge}}=e/2$ charges. We consider the case where the edge is metallic and is half filled in accordance with the edge charge. (b) After the surface reconstruction, the edge periodicity is doubled ($a^{\prime}_{\text{edge}}=2a_{\text{edge}})$. Then, each edge unit cell contains integer charges, and the edge can be gapped and insulating. 
}
    \label{surface_reconstruction}
\end{figure}
Here we consider systems with surface reconstruction. Conceptual pictures of the surface reconstruction is shown in Fig.~\ref{surface_reconstruction}. We assume that the system has nonzero quantized electric polarization in the bulk. Then the edge have nonzero fractional edge charge $\sigma_{\text{edge}}$. In Fig.~\ref{surface_reconstruction}(a) we show its band structure when the edge states are metallic. However, it can under go Peierls transition or surface reconstruction to multiply edge periodicity and open a gap. In Fig.~\ref{surface_reconstruction}(b), due to the surface reconstruction, the surface periodicity is doubled, and the edge charge becomes an integer in the enlarged surface unit cells. Then, the edge energy spectrum can be gapped and insulating. In this case, one can apply our general formula.

\subsection{General edge orientations}
\begin{figure}
  \centerline{\includegraphics[width=8cm,clip]{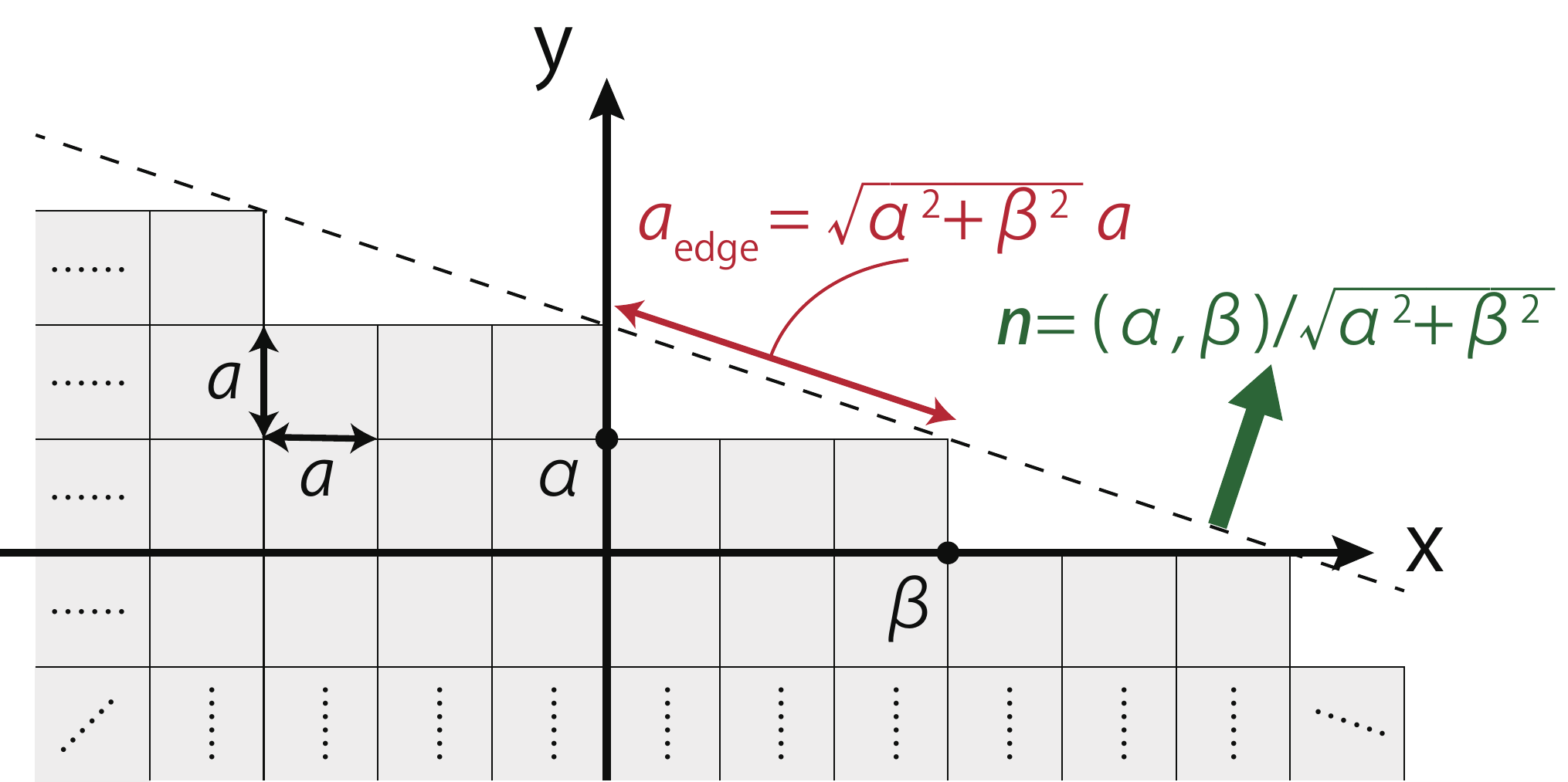}}
  \caption{(Color online) Conceptual picture of the surface with high Miller index ($\alpha,\beta$). The gray squares are unit cells. 
}
    \label{high_Miller_index}
\end{figure}

Here, we consider the surface with a high Miller index. First, we consider the $C_4$-symmetric system with basis vectors of $\bm{a}_1=(a,0)$ and $\bm{a}_2=(0,a)$, as shown in Fig.~{\ref{Cn_Wyckoff}}(b). 
In the $C_4$-symmetric system, the bulk electric polarization is quantized to be 
$\bm{P}\equiv(0,0)$ or $(\frac{e}{2a},\frac{e}{2a})$ (mod $\frac{e}{a}(m_1,m_2)$), ($m_1,m_2\in\mathbb{Z}$)\cite{PhysRevB.99.245151,PhysRevB.102.165120,PhysRevB.48.4442}. 
In the following, we consider the case with $\bm{P}\equiv(\frac{e}{2a},\frac{e}{2a})$.
Next, for the surface with a Miller index ($\alpha,\beta$) ($\alpha,\beta\in\mathbb{Z}$, and $\alpha$ and $\beta$ are coprime), the surface normal vector $\bm{n}$ is defined as follows: 
\begin{align}
\bm{n}=\Bigg{(}
\frac{\alpha}{\sqrt{\alpha^2+\beta^2}},\ 
\frac{\beta}{\sqrt{\alpha^2+\beta^2}}
\Bigg{)}.
\end{align}
Therefore, the edge charge in one edge unit cell is calculated as follows: 
\begin{align}
\sigma a_{\text{edge}}
\equiv\bm{P}\cdot\bm{n} a_{\text{edge}}
\equiv\frac{e}{2}(\alpha+\beta)\ (\text{mod}\ e).
\end{align}
Here, we set the size of the edge unit cell to be the minimal one, $a_{\text{edge}}=|\beta\bm{a}_1-\alpha\bm{a}_2|=\sqrt{\alpha^2+\beta^2}a$, as shown in Fig.~\ref{high_Miller_index}. 
If both $\alpha$ and $\beta$ are odd integers, $\sigma a_{\text{edge}}\equiv0$ (mod $e$), and otherwise $\sigma a_{\text{edge}}\equiv e/2$ (mod $e$). Obviously, the edge can be charge neutral when $\sigma a_{\text{edge}}\equiv0$ (mod $e$). Even when $\sigma a_{\text{edge}}\equiv e/2$ (mod $e$), after surface reconstruction with doubling the surface period $a_{\text{edge}}^{\prime}=2a_{\text{edge}}$, we get $\sigma a_{\text{edge}}^{\prime}\equiv0$ (mod $e$), and the edge can be charge neutral. In particular, when the system has edge states in the gap, and the fractional edge charge is accommodated in the edge states, the edge states are half-filled and it always has Peierls instability, meaning that the edge becomes insulating via spontaneous lattice deformation with doubling the edge unit cell at sufficiently low temperature.

Next, we consider the $C_3$-symmetric system with basis vectors of $\bm{a}_1=(\frac{a}{2},\frac{\sqrt{3}}{2}a)$ and $\bm{a}_2=(-\frac{a}{2},\frac{\sqrt{3}}{2}a)$, as shown in Fig.~{\ref{Cn_Wyckoff}}(a). In the $C_3$-symmetric system, the bulk electric polarization is $\bm{P}=\bm{0}$, $\frac{e}{3\Omega}(\bm{a}_1+\bm{a}_2)$ or $\frac{-e}{3\Omega}(\bm{a}_1+\bm{a}_2)$ (mod $\frac{e}{\Omega}\bm{R}$)\cite{PhysRevB.99.245151,PhysRevB.102.165120,PhysRevB.48.4442}, where $\Omega=\frac{\sqrt{3}}{2}a^2$ is the area of the bulk unit cell and $\bm{R}=m_1\bm{a}_1+m_2\bm{a}_2$ ($m_1,m_2\in\mathbb{Z}$) is a lattice vector. In the following, we consider the case with $\bm{P}=\frac{\pm e}{3\Omega}(\bm{a}_1+\bm{a}_2)=(0,\frac{\pm2e}{3a})$. The surface normal vector $\bm{n}$ for the surface with Miller index ($\alpha,\beta$) ($\alpha,\beta\in\mathbb{Z}$, and $\alpha$ and $\beta$ are coprime) is defined as follows:
\begin{align}
\bm{n}
&=\frac{1}{\sqrt{\alpha^2-\alpha\beta+\beta^2}}
\Bigg{(}
\frac{\sqrt{3}}{2}(\alpha-\beta),\ 
\frac{1}{2}(\alpha+\beta)
\Bigg{)}. 
\end{align}
Therefore, the edge charge in one edge unit cell is calculated as follows: 
\begin{align}
\sigma a_{\text{edge}}
\equiv\bm{P}\cdot\bm{n} a_{\text{edge}}
\equiv\pm\frac{e}{3}(\alpha+\beta)\ (\text{mod}\ e).
\end{align}
Here, we used $a_{\text{edge}}=|\beta\bm{a}_1-\alpha\bm{a}_2|=\sqrt{\alpha^2-\alpha\beta+\beta^2}a$. 
If $\alpha+\beta\in3\mathbb{Z}$, $\sigma a_{\text{edge}}\equiv0$ (mod $e$), and otherwise $\sigma a_{\text{edge}}\equiv e/3$ or $(2e)/3$ (mod $e$). Obviously, edge can be charge neutral when $\sigma a_{\text{edge}}\equiv0$ (mod $e$). Even when $\sigma a_{\text{edge}}\equiv e/3$ or $(2e)/3$ (mod $e$), after surface reconstruction with tripling the surface period $a_{\text{edge}}^{\prime\prime}=3a_{\text{edge}}$, we get $\sigma a_{\text{edge}}^{\prime\prime}\equiv0$ (mod $e$), and the edge can be charge neutral.

Finally, we consider the $C_6$-symmetric system with basis vectors of $\bm{a}_1=(a,0)$ and $\bm{a}_2=(-\frac{a}{2},\frac{\sqrt{3}}{2}a)$, as shown in Fig.~{\ref{Cn_Wyckoff}}(c). In the $C_6$-symmetric system, the bulk electric polarization is always $\bm{P}=\bm{0}$ (mod $\frac{e}{\Omega}\bm{R}$)\cite{PhysRevB.99.245151,PhysRevB.102.165120,PhysRevB.48.4442}, where $\Omega=\frac{\sqrt{3}}{2}a^2$ is the area of the bulk unit cell and $\bm{R}=(m_1a-\frac{m_2}{2}a,\frac{\sqrt{3}}{2}m_2a)$ ($m_1,m_2\in\mathbb{Z}$) is a lattice vector. 
Therefore, $\sigma_{\text{edge}}=\bm{P}\cdot\bm{n}$ is 0 for surfaces in any direction (mod $e/a_{\text{edge}}$).

To summarize, in either case, even when $\bm{P}\neq0$ in the $C_n$-symmetric systems, $\sigma a_{\text{edge}}\equiv0$ (mod $e$) can be achieved with proper surface reconstruction thanks to the fractional quantization of $\bm{P}$.

\section{Application to various systems}

\subsection{Tight-binding models}
Here, we will explain how to apply our formula to tight-binding models. In principle, tight-binding models contain only electron degrees of freedom. 
In order to calculate the corner charge, it is necessary to give appropriate information on the positions of ions.

First, it should be pointed out that the sites of the tight-binding models are different from the positions of ions. 
The apatite electride is one of the examples showing that these two are independent \cite{PhysRevResearch.2.043131}. 
The electronic states near the Fermi level of the apatite electride are well described by the Wannier functions localized at the one-dimensional hollows in the crystal. Therefore the tight-binding model made from these Wannier orbitals has its sites inside the hollows. 
Since there are no ions in the hollows, the ion positions and the sites are clearly different in this model.

Next, we apply our formula to the well-known 2D Benalcazar-Bernevig-Hughes (BBH) model \cite{Benalcazar61,PhysRevB.96.245115} as an example. 
This model is a spinless tight-binding model on a square lattice with $\pi$-flux, showing a higher-order topological insulating phase. 
As in Ref.~[\onlinecite{PhysRevB.96.245115}], we assume that all the positive ions are located at the center of the $C_4$-symmetric unit cell, i.e., $n_{1a}^{(\text{ion})}=\nu$ holds. 
The $k$-space Hamiltonian is given as follows:
\begin{align}
\mathcal{H}(\bm{k})&=(\gamma+\lambda\cos k_x)\tau_x\sigma_0-\lambda\sin k_x\tau_y\sigma_z 
\notag \\
&\quad-(\gamma+\lambda\cos k_y)\tau_y\sigma_y-\lambda\sin k_y\tau_y\sigma_x. 
\end{align}
Here, $\gamma$ and $\lambda$ are real parameters, and $\tau_j$, $\sigma_j$ ($j=x,y,z$) are Pauli matrices. 
For simplicity, we take the lattice constants as $a=1$. 
The Hamiltonian has $C_4$-symmetry $\hat{C_4}$:
\begin{align}
\hat{C_4}&=
\begin{pmatrix}
0 & \sigma_0 \\
-i\sigma_y & 0
\end{pmatrix}_{\tau}.
\end{align}
Here, due to the $\pi$-flux, $\hat{C_4}$ satisfies $(\hat{C_4})^4=-1$, and its eigenvalues are $C_4=e^{\frac{(2n+1)\pi i}{4}}$ ($n=0,1,2,3$). 
In the derivation of the corner charge formula in this paper, we assumed that $(\hat{C_n})^n=1$ for spinless systems. 
Therefore, our formula cannot be applied as it is, when the rotation eigenvalues are modified due to the $\pi$-flux.
However, we can formally determine the corner charge by applying the formula for the spinful class A system as if the model described a spinful system because $(\hat{C_4})^4=-1$. 


For simplicity, we consider two limits: $\lambda=0$ and $\gamma=0$. 
When $\lambda=0$, the Hamiltonian becomes $\mathcal{H}(\bm{k})=\gamma(\tau_x\sigma_0-\tau_y\sigma_y)$, which does not depend on $k$. Then topological invariants are trivial: $[X_1^{(2)}]=[M_1^{(4)}]=[M_2^{(4)}]=0$. 
When $\gamma=0$, the Hamiltonian becomes
\begin{align}
\mathcal{H}(\bm{k})
&=\sqrt{2}\lambda
\begin{pmatrix}
0 & Q^{\dagger}(\bm{k}) \\
Q(\bm{k}) & 0
\end{pmatrix}, 
\\
Q(\bm{k})&=
\frac{1}{\sqrt{2}}
\begin{pmatrix}
e^{-ik_x} & -e^{ik_y} \\
e^{-ik_y} & e^{ik_x}
\end{pmatrix}.
\end{align}
Topological invariants can be calculated by taking the simultaneous eigenstates of the Hamiltonian and the rotation operation $\hat{C_4}$ or $\hat{C_2}=(\hat{C_4})^2$ at each of the high symmetry points $\Gamma=(0,0)$, $X=(\pi,0)$, and $M=(\pi,\pi)$. 
The results are summarized in Table.~\ref{tab:BBH}, and the corner charge is calculated from the Eq.(\ref{Qc_4_A}):
\begin{align}
Q_{c\_1a}^{(4)}
\equiv\frac{|e|}{4}\Big{(}2-2+0-\frac{1}{2}\cdot1+\frac{3}{2}\cdot(-1)\Big{)}
\equiv \frac{e}{2}\ (\text{mod}\ e).
\end{align}
This result agrees with the results of previous studies\cite{Benalcazar61,PhysRevB.96.245115}. 
This value of the corner charge is stable as long as $|\gamma|<|\lambda|$, where the bulk gap is open.

\begin{table}
\begin{tabular}{|c|c|c|c|c|c|c|c|}
		\hline
Model & $n_{1a}^{(\text{ion})}$ & $\nu$ & $[X^{(2)}_1]$ & $[M_1^{(4)}]$ & $[M_2^{(4)}]$ & $Q_{c\_1a}=Q_{c\_1b}$ (mod $e$)  \\
		\hline
 BBH & 2 & 2 & 0 & 1 & $-1$ & $e/2$ (mod $e$)  \\
	\hline
\end{tabular}
\caption{Topological invariants and a corner charge of the BBH model with $|\gamma|<|\lambda|$. 
}
	\label{tab:BBH}
\end{table}

\subsection{First-principles calculations}

\begin{figure}
  \centerline{\includegraphics[width=8cm,clip]{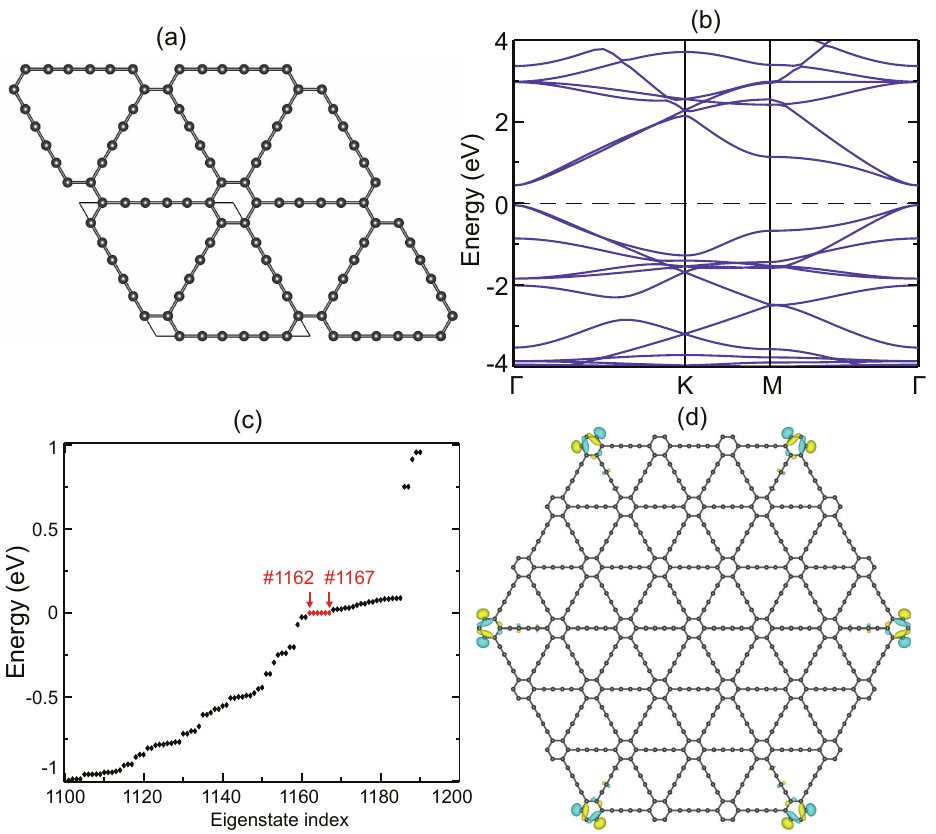}}
  \caption{(Color online) First-principles calculations for graphdiyne. (a) Crystal structure. The circles denote carbon atoms. (b) Spinless electronic band structure. The Fermi energy is in the middle of the gap, at about $0.2$ eV. (c) Low-energy spectrum of a finite $C_6$-preserved nanodisk. The six red dots represent corner localized states. When the state below $\#1164$ is occupied, the system is charge neutral. (d) Charge distributions of the $\#1162$ state located at six corners.} 
\label{DFT_Graphydine}
\end{figure}

\begin{table}
\begin{tabular}{|c|c|c|c|c|c|}
		\hline
Material & $n_{1a}^{(\text{ion})}$ & $\nu$ & $[M_1^{(2)}]$ & $[K^{(3)}_1]$ & $Q_{c\_1a}$ (mod $e$) \\
		\hline
	Graphdiyne & 0 & 36 &	 $-2$ & 0 & $e/2$ (mod $e$) \\
	\hline
\end{tabular}
\caption{Topological invariants and a corner charge of the graphdiyne. }
	\label{tab:Graphydine}
\end{table}

In this subsection, we explain $ab\ initio$ calculations for two materials, to show agreement with our corner-charge formula.

The first example is graphdiyne, which was studied in Ref.~[\onlinecite{PhysRevLett.123.256402}]
as a $C_6$-symmetric higher-order topological insulator with a fractional corner charge. 	Figure~\ref{DFT_Graphydine}(a) shows the crystal structure of graphdiyne, which belongs to the wallpaper group $p6mm$
with $C_6$ rotation symmetry. Since the spin-orbit coupling does not lead to additional band inversion in graphdiyne and therefore does not influence the corner charge, we calculate the spinless electronic band structure of graphdiyne, which is shown in Fig.~\ref{DFT_Graphydine}(b), after a full relaxation on atomic positions until the forces on the ions are less than 0.01 eV/\AA\ via Vienna $ab\ initio$ simulation package (VASP) \cite{PhysRevLett.107.107403,Kogar1314,lee1993conductivity}. An $11\times11\times1$ $k$-mesh is used in the BZ for the self-consistent calculations with a plane-wave energy cutoff of 500 eV. In a $C_6$-preserved nanodisk configuration, the eigenvalues are shown in Fig.~\ref{DFT_Graphydine}(c), which have six degenerate states $\#1162$, $\cdots$ $\#1167$ around the Fermi energy (marked by red dots). These six states are localized at six corners, as is seen in Fig.~\ref{DFT_Graphydine}(d) showing the charge distributions of the $\#1162$ state. 
Because the number of electrons is $1164$, we need to add or subtract three electrons to make these six states fully occupied or empty to make the system gapped. Therefore, the value of the corner charge is $\pm\frac{3e}{6}=\frac{e}{2}$ (mod $e$).
This is consistent with the corner charge calculation from our formula in Tab.~\ref{tab:Graphydine}.

\begin{figure}
  \centerline{\includegraphics[width=8cm,clip]{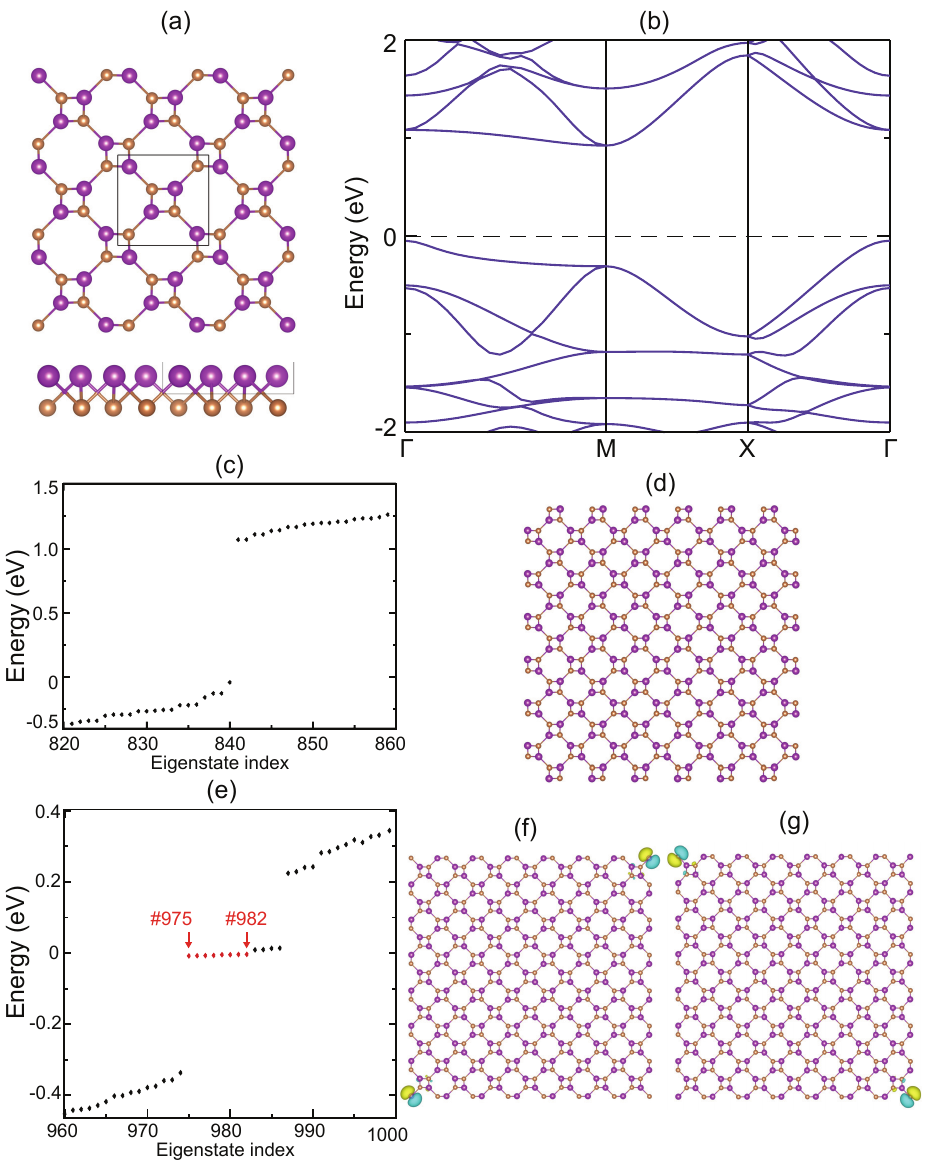}}
  \caption{(Color online) First-principles calculations for BiSb. (a) Crystal structure. The purple and orange circles denote Bi and Sb atoms, respectively. (b) Spinless electronic band structure. The Fermi energy is in the middle of the gap, at about $0.5$ eV. (c) Low-energy spectrum of a finite $C_4$-preserved nanodisk shown in (d). The number of electrons is $840$, and the spectrum is gapped at $E_F$. 
(e) Low-energy spectrum of a finite $C_2$-preserved nanodisk. The twelve red dots represent corner localized states. When the state below $\#980$ is occupied, the system is charge neutral. (f-g) Charge distributions of the (f) $\#980$ and (g) $\#981$ states.} 
\label{DFT_BiSb}
\end{figure}

\begin{table}
\begin{tabular}{|c|c|c|c|c|c|c|}
		\hline
Material & $n_{1a}^{(\text{ion})}$ & $\nu$ & $[X^{(2)}_1]$ & $[M_1^{(4)}]$ & $[M_2^{(4)}]$ & $Q_{c\_1a}$ (mod $e$) \\
		\hline
	BiSb & 0 & 20 & $-2$ & 8 & 2 & $0$ (mod $e$) \\
	\hline
\end{tabular}
\caption{Topological invariants and a corner charge of BiSb. 
}
	\label{tab:BiSb}
\end{table}

The other example is $C_4$-symmetric two-dimensional BiSb. 
Figure~\ref{DFT_BiSb}(a) show both the top and side views for the crystal structure of BiSb, which belongs to the wallpaper group $p4gm$ 
with $C_4$ rotation symmetry. Figure~\ref{DFT_BiSb}(b) is the spinless electronic band structure also calculated by VASP after a full relaxation on atomic positions. An $8\times8\times1$ $k$-mesh is used in the BZ for the self-consistent calculations with a plane-wave energy cutoff of 320 eV.
From the corner charge formula calculation in Tab.~\ref{tab:BiSb}, there is no corner charge in a $C_4$-preserved nanodisk of BiSb, as shown in Fig.~\ref{DFT_BiSb}(c-d). 
Namely, the gap is open everywhere including the corners. It means that the filling anomaly is zero. 
However, the corner state can exist in a $C_4$-breaking configuration. Figure~\ref{DFT_BiSb}(e) shows the eigenstates of a $C_4$-broken $C_2$-preserved nanodisk, which have 8 states around the Fermi energy. Figures~\ref{DFT_BiSb}(f-g) are the charge distributions of the $\#980$ and $\#981$ state.
The number of electrons is $980$. Therefore, corner states appears, but because $C_4$-symmetry is not preserved, the corner charge is not quantized. 
These corner states are not symmetry-protected and their appearance depends on the surface termination.
Their appearance is attributed to dangling bonds at the corners.


\section{Conclusion and Discussion}
In the present paper, we derived the general formula for the quantized fractional corner charge under the assumptions that the electronic states can be described by Wannier orbitals and the edges are charge neutral. These assumptions are more general than the previous studies, which assumed vanishing bulk electric polarization, and these formulas are more general to be used in different systems with/without time-reversal symmetry/spin-orbit coupling. 
We expanded the scope of the corner charge formula by considering more general surface conditions, such as surfaces with higher Miller index and surfaces with surface reconstruction.

Our proof also shows that even when the positions of the nuclei and the electronic states are largely modulated near the boundaries, the corner charge is still quantized to fractional values as long as the system is $C_n$-symmetric and the edge charge density is zero. Thus, unlike previous theories, our theory also includes some insulators with non-vanishing bulk polarization, and in such cases with $C_3$ or $C_4$ symmetries, we find that the value of the corner charge depends on the Wyckoff position of the center of the $C_n$-symmetric crystal. 

We note that the corner charge formula has a simple form like Eq.~(\ref{Q_c_general}) only when the system has $C_n$ rotational symmetry. The corner charge in systems without rotational symmetry is discussed in Refs.~[\onlinecite{PhysRevResearch.2.043012}, \onlinecite{PhysRevB.103.035147}].

Finally, we briefly comment on the generalization of our results to three-dimensional systems. 
There are two directions for generalization: hinge charge calculation and corner charge calculation.
First, let us consider a case where a three-dimensional system can be regarded as a stack of $C_n$-symmetric two-dimensional systems, and cut the system into $n$-gonal prism. 
Then, the hinge charge per unit length is equal to the corner charge of the two-dimensional layer divided by the hinge period $a_{\text{hinge}}$. 
In this case, our formula can be used to determine the hinge charge. 
However, it is not clear that our formula can be generalized to the calculation of the hinge charge of general three-dimensional systems, and it is future work. 

The generalization to the calculation of corner charges is a more non-trivial problem. 
In order to discuss corner charges in three-dimensional systems, it is necessary to discuss the charge neutrality of the hinge in addition to the charge neutrality of the bulk and the surface. However, it is not clear whether the charge neutrality of the hinge can be determined only from the bulk quantity. 
Furthermore, three-dimensional systems can be cut out in various ways, and the same crystal may have different corner charges. 
For example, in the case cubic symmetry, both octahedron and cube maintain the cubic symmetry. 
These two may have different corner charges due to the different number of corners. 

\textit{Note added.} A recent publication \cite{PhysRevB.103.165109}
computes the filling anomaly for  two- and three-dimensional $C_4$ symmetric lattices, and our results agree with
it for two-dimensional $C_4$ symmetric lattices.

\begin{acknowledgments}
R.T. thanks K. Naitou for useful comments. 
This work was supported by JSPS KAKENHI Grant Numbers JP18J23289, JP18H03678, and JP20H04633. 
\end{acknowledgments}

\appendix{}
\section{Relationship between the topological invariants $[\Pi_{p}^{(n)}]$ and $n^{(e)}_{mX}$ in class A systems}

Here, we discuss the relationship between the topological invariants $[\Pi_{p}^{(n)}]$ and the occupation number for Wyckoff positions $n^{(e)}_{mX}$. 
We use a technique similar to that used in Refs.~[\onlinecite{PhysRevB.99.245151}, \onlinecite{PhysRevResearch.1.033074}, \onlinecite{PhysRevResearch.2.043131}], as we explain in detail in the following. 
The topological invariant $[\Pi_{p}^{(n)}]$ represents the difference of the number of states with $C_n$ eigenvalue $e^{\frac{2(p-1)\pi i}{n}}$ for spinless systems, and $e^{\frac{(2p-1)\pi i}{n}}$ for spinful systems, between the rotation-invariant $\bm{k}$-points $\Pi$ and $\Gamma$.
First, we assume that the eigenstates can be described by well-localized Wannier orbitals. Thanks to the discussion based on lattice homotopy, we can limit ourselves to Wannier orbitals localized at one of the high-symmetry Wyckoff positions\cite{PhysRevLett.119.127202}. 
Thus, we calculate the topological invariants for such Wannier configurations $W_j$\cite{PhysRevB.99.245151,PhysRevResearch.1.033074}, and Tables~\ref{tab:PrimitiveGenerators_C3}, \ref{tab:PrimitiveGenerators_C4} and  \ref{tab:PrimitiveGenerators_C6} summarize the values of the topological number $[\Pi_{p}^{(n)}]$ for Wannier configurations for $C_3$-, $C_4$- and $C_6$-symmetric systems, respectively. 
Different from the Refs.~[\onlinecite{PhysRevB.99.245151},\onlinecite{PhysRevResearch.1.033074}], we calculate the general corner charge formula including the case where the ion position is not limited to $1a$ Wyckoff positions. 
Since the eigenstates are expressed in terms of Wannier orbitals, the topological invariant $\bm{\chi}$ for the occupied bands is represented as the summation of the topological invariants of the Wannier configurations $\bm{\chi}^{(j)}$:
\begin{align}
\bm{\chi}&=\sum_{j=1}^{M} \alpha_j\bm{\chi}^{(j)}\quad (\alpha_j\in\mathbb{Z}_{\geqslant}), 
\\
&=C_{\bm{\chi}}\bm{\alpha}.
\end{align}
Here, $M$ is the number of Wannier configurations with nonzero topological invariants, $C_{\bm{\chi}}=(\bm{\chi}^{(1)},\bm{\chi}^{(2)},\cdots,\bm{\chi}^{(M)})$ and $\bm{\alpha}=(\alpha_1,\alpha_2,\cdots,\alpha_M)^{\text{T}}$, which represents the numbers of occupied bands corresponding to the Wannier configurations $W_j$. 
By definition, $n^{(e)}_{mX}$ is represented by the sum of $\alpha_j$ belonging to the Wyckoff  position $mX$. 
Since $C_{\bm{\chi}}$ is generally not a square matrix and is not always invertible, $\alpha_j$ itself cannot be uniquely determined from $\bm{\chi}$. However, in a $C_N$ symmetric system, $n^{(e)}_{mX}$, expressed in terms of $\alpha_j$, can be determined modulo $N$ ($N=3,4,6$) through direct calculations, which is sufficient for the calculation of the corner charge.

\begin{figure}[t]
  \centerline{\includegraphics[width=8cm,clip]{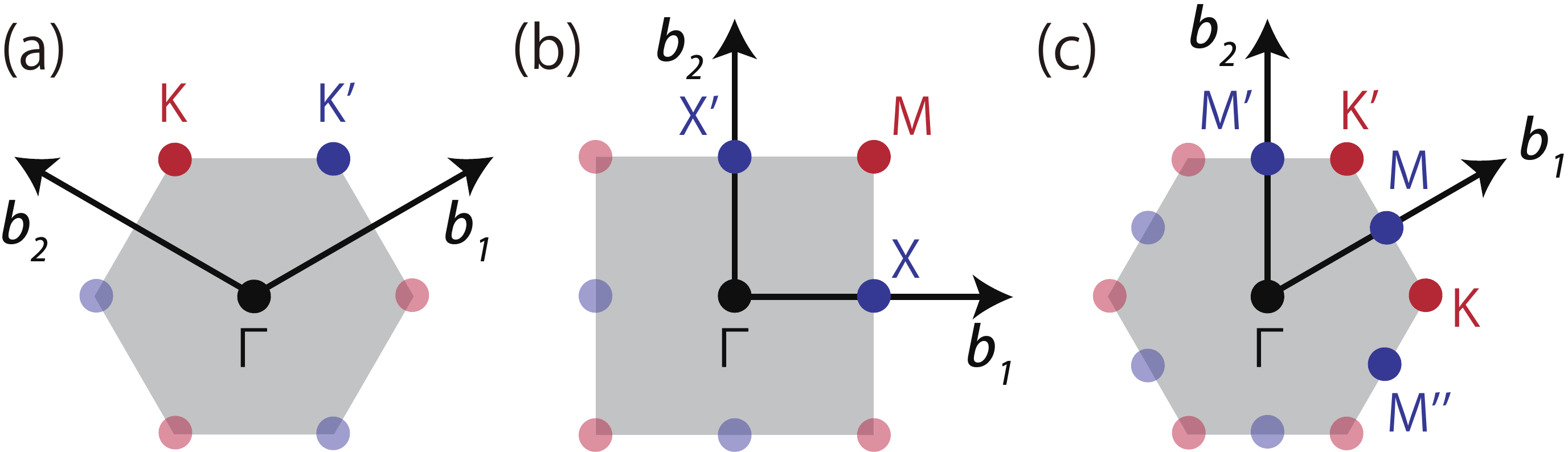}}
  \caption{(Color online)  Highly-symmetric $\bm{k}$-points in (a) $C_3$, (b) $C_4$, (c) $C_6$-symmetric systems. The points with same colors belong to same $\bm{k}$-vector star, and related by $C_n$-operation. Here, $\bm{b}_1$ and $\bm{b}_2$ are reciprocal lattice vector, and (a) $\bm{b}_1=2\pi(\frac{\sqrt{3}}{2},\frac{1}{2})\frac{2}{\sqrt{3}a}$, $\bm{b}_2=2\pi(-\frac{\sqrt{3}}{2},\frac{1}{2})\frac{2}{\sqrt{3}a}$, (b) $\bm{b}_1=(1,0)\frac{2\pi}{a}$, $\bm{b}_2=(0,1)\frac{2\pi}{a}$, (c) $\bm{b}_1=2\pi(\frac{\sqrt{3}}{2},\frac{1}{2})\frac{2}{\sqrt{3}a}$, $\bm{b}_2=2\pi(0,1)\frac{2}{\sqrt{3}a}$. 
}
    \label{Cn_BZ}
\end{figure}

\subsection{$C_3$-symmetry}
\begin{table}
\begin{tabular}{cccccc}
		\hline
		\hline
Label & Wannier configurations & \multicolumn{4}{c}{invariants}\\
		\hline
		\hline
	$j$ &	   & $[K^{(3)}_1]$ & $[K^{(3)}_2]$ & $[K^{\prime(3)}_1]$ & $[K^{\prime(3)}_2]$\\
		\hline
1&$W_{1b|_{0}}^{(3)}$ & $-1$ & 0 & $-1$ & 1 \\
2&$W_{1b|_{2\pi/3}}^{(3)}$ & 1 & $-1$ & 0 & $-1$ \\
3&$W_{1b|_{4\pi/3}}^{(3)}$ & 0 & 1 & 1 & 0 \\
4&$W_{1c|_{0}}^{(3)}$ & $-1$ & 1 & $-1$ & 0 \\
5&$W_{1c|_{2\pi/3}}^{(3)}$ & 0 & $-1$ & 1 & $-1$ \\
6&$W_{1c|_{4\pi/3}}^{(3)}$ & 1 & 0 & 0 & 1 \\
	\hline
	\hline
\end{tabular}
\caption{
Wannier configurations with labels and their invariants with $C_3$ symmetry in class A systems. 
The first column is the label for the WCs. 
The Wannier configurations belonging to Wyckoff positions $1a$ and $3d$ are omitted in the table because all topological invariants are 0.}
	\label{tab:PrimitiveGenerators_C3}
\end{table}

In $C_3$-symmetric systems, we need to consider six Wannier configurations $W_j$ ($j=1,2,\cdots,6$) in Table~\ref{tab:PrimitiveGenerators_C3}. For example, $W_1=W_{1b|_{0}}^{(3)}$ represents a Wannier configuration localized at Wyckoff position $1b$ with a $C_3$-eigenvalue of $e^{i\times0}$. Likewise, $W_{mX|_{\theta}}^{(n)}$ denotes a Wannier configuration localized at Wyckoff position $mX$ with a $C_n$-eigenvalue of $e^{i\theta}$. 
Table~\ref{tab:PrimitiveGenerators_C3} is summarized as follows:
\begin{align}
\begin{pmatrix}
[K^{(3)}_1] \\
[K^{(3)}_2] \\
[K^{\prime(3)}_1] \\
[K^{\prime(3)}_2] 
\end{pmatrix}
=
\begin{pmatrix}
-1 & 1 & 0 & -1 & 0 & 1 \\
0 & -1 & 1 & 1 & -1 & 0 \\
-1 & 0 & 1 & -1 & 1 & 0 \\
1 & -1 & 0 & 0 & -1 & 1 
\end{pmatrix}
\begin{pmatrix}
\alpha_1 \\
\alpha_2 \\
\vdots \\
\alpha_6 
\end{pmatrix}. 
\label{C3_WC_invariants}
\end{align}
Here, $K=\frac{2}{3}\bm{b}_1+\frac{1}{3}\bm{b}_2$ and $K^{\prime}=\frac{1}{3}\bm{b}_1+\frac{2}{3}\bm{b}_2$ (see Fig.~{\ref{Cn_BZ}}(a)). 
From the definition, $n^{(e)}_{1b}=\alpha_1\!+\!\alpha_2\!+\!\alpha_3$ and $n^{(e)}_{1c}=\alpha_4\!+\!\alpha_5\!+\!\alpha_6$. 
From Eq.~(\ref{C3_WC_invariants}), the following relation holds:
\begin{align}
[K^{(3)}_2] +[K^{\prime(3)}_1] 
&=-\alpha_1-\alpha_2+2\alpha_3
\notag \\
&\equiv-(\alpha_1+\alpha_2+\alpha_3)\ (\text{mod}\ 3),
\\
[K^{(3)}_1] +[K^{\prime(3)}_2]
&=-\alpha_4-\alpha_5+2\alpha_6
\notag \\
&\equiv-(\alpha_4+\alpha_5+\alpha_6)\ (\text{mod}\ 3).
\end{align}
Therefore, $n^{(e)}_{1b}$ and $n^{(e)}_{1c}$ (mod $3$) can be determined from $[\Pi_{p}^{(n)}]$ as follows:
\begin{align}
&n^{(e)}_{1b}=-[K_2^{(3)}]-[K_1^{\prime(3)}]
\ (\text{mod}\ 3), 
\\
&n^{(e)}_{1c}=-[K_1^{(3)}]-[K_2^{\prime(3)}]
\ (\text{mod}\ 3). 
\end{align}
Then, $n^{(e)}_{1a}=\nu-n^{(e)}_{1b}-n^{(e)}_{1c}-3n^{(e)}_{3d}\equiv\nu-n^{(e)}_{1b}-n^{(e)}_{1c}$ (mod 3) is also determined from $\nu$ and $[\Pi_{p}^{(n)}]$. 
Finally, the corner charge formulas are represented as follows: 
\begin{align}
Q_{c\_1a}^{(3)}
&\equiv\frac{|e|}{3}\Big{(}n_{1a}^{(\text{ion})}-\nu-[K_1^{(3)}]-[K_2^{(3)}]-[K_1^{\prime(3)}]-[K_2^{\prime(3)}]\Big{)} 
\notag \\
&\ \quad\quad\quad\quad\quad\quad\quad\quad\quad\quad\quad\quad\quad\quad\ (\text{mod}\ e),
\\
Q_{c\_1b}^{(3)}
&\equiv\frac{|e|}{3}\Big{(}n_{1b}^{(\text{ion})}+[K_2^{(3)}]+[K_1^{\prime(3)}]
\Big{)}\ (\text{mod}\ e),
\\
Q_{c\_1c}^{(3)}
&\equiv\frac{|e|}{3}\Big{(}n_{1c}^{(\text{ion})}+[K_1^{(3)}]+[K_2^{\prime(3)}]
\Big{)}\ (\text{mod}\ e).
\end{align}

\subsection{$C_4$-symmetry}
\begin{table}
\begin{tabular}{cccccc}\hline\hline
Label & Wannier configurations & \multicolumn{4}{c}{invariants}\\
\hline
\hline
	$j$	   & & $[X^{(2)}_1]$ & $[M^{(4)}_1]$ & $[M^{(4)}_2]$ & $[M^{(4)}_3]$\\
\hline
1&$W_{1b|_{0}}^{(4)}$ & $-1$ & $-1$ & 0 & 1 \\
2&$W_{1b|_{\pi/2}}^{(4)}$ & 1 & 0 & $-1$ & 0 \\
3&$W_{1b|_{\pi}}^{(4)}$ & $-1$ & 1 & 0 & $-1$ \\
4&$W_{1b|_{3\pi/2}}^{(4)}$ & 1 & 0 & 1 & 0 \\
5&$W_{2c|_{0}}^{(4)}$ & $-1$ & $-1$ & 1 & $-1$ \\
6&$W_{2c|_{\pi}}^{(4)}$ & 1 & 1 & $-1$ & 1 \\
	\hline
	\hline
\end{tabular}
\caption{Wannier configurations with labels and their invariants with $C_4$ symmetry in class A systems. 
The first column is the label for the WCs. 
The Wannier configurations belonging to Wyckoff positions $1a$ and $4d$ are omitted in the table because all topological invariants are 0.
}
	\label{tab:PrimitiveGenerators_C4}
\end{table}

In $C_4$-symmetric systems, we need to consider six Wannier configurations in Table~\ref{tab:PrimitiveGenerators_C4}. 
Table~\ref{tab:PrimitiveGenerators_C4} is summarized as follows:
\begin{align}
\begin{pmatrix}
[X^{(2)}_1] \\
[M^{(4)}_1] \\
[M^{(4)}_2] \\
[M^{(4)}_3] 
\end{pmatrix}
=
\begin{pmatrix}
-1 & 1 & -1 & 1 & -1 & 1 \\
-1 & 0 & 1 & 0 & -1 & 1 \\
0 & -1 & 0 & 1 & 1 & -1 \\
1 & 0 & -1 & 0 & -1 & 1 
\end{pmatrix}
\begin{pmatrix}
\alpha_1 \\
\alpha_2 \\
\vdots \\
\alpha_6 
\end{pmatrix}. 
\label{C4_WC_invariants}
\end{align}
Here, $X=\frac{1}{2}\bm{b}_1$ and $M=\frac{1}{2}\bm{b}_1+\frac{1}{2}\bm{b}_2$ (see Fig.~{\ref{Cn_BZ}}(b)). 
From the definition, $n^{(e)}_{1b}=\alpha_1\!+\!\alpha_2\!+\!\alpha_3\!+\!\alpha_4$ and $n^{(e)}_{2c}=\alpha_5\!+\!\alpha_6$. 
From Eq.~(\ref{C4_WC_invariants}), the following relation holds:
\begin{align}
[X^{(2)}_1]\!-\!\frac{3}{2}[M^{(4)}_1]\!+\!\frac{1}{2}[M^{(4)}_3]
&=\alpha_1+\alpha_2-3\alpha_3+\alpha_4
\notag \\
&\equiv \alpha_1+\alpha_2+\alpha_3+\alpha_4 \ (\text{mod}\ 4),
\\
[M^{(4)}_1] + [M^{(4)}_3]
&=-2\alpha_5+2\alpha_6
\notag \\
&\equiv 2(\alpha_5+\alpha_6) \ (\text{mod}\ 4).
\end{align}
Therefore, $n^{(e)}_{1b}$ (mod $4$) and $n^{(e)}_{2c}$ (mod $2$) can be determined from $[\Pi_{p}^{(n)}]$ as follows:
\begin{align}
&n^{(e)}_{1b}=[X_1^{(2)}]-\frac{3}{2}[M_1^{(4)}]+\frac{1}{2}[M_3^{(4)}]
\ (\text{mod}\ 4), 
\\
&n^{(e)}_{2c}=\frac{1}{2}[M_1^{(4)}]+\frac{1}{2}[M_3^{(4)}]
\ (\text{mod}\ 2). 
\end{align}
Then, $n^{(e)}_{1a}=\nu-n^{(e)}_{1b}-2n^{(e)}_{2c}-4n^{(e)}_{4d}\equiv\nu-n^{(e)}_{1b}-2n^{(e)}_{2c}$ (mod 4) is also determined from $\nu$ and $[\Pi_{p}^{(n)}]$. 
Finally, the corner charge formulas are summarized as follows: 
\begin{align}
Q_{c\_1a}^{(4)}
&\equiv\frac{|e|}{4}\Big{(}n_{1a}^{(\text{ion})}-\nu+[X_1^{(2)}]-\frac{1}{2}[M_1^{(4)}]+\frac{3}{2}[M_3^{(4)}]
\Big{)} 
\notag \\
&\ \quad\quad\quad\quad\quad\quad\quad\quad\quad\quad\quad\quad\quad\quad\ (\text{mod}\ e),
\\
Q_{c\_1b}^{(4)}
&\equiv\frac{|e|}{4}\Big{(}n_{1b}^{(\text{ion})}-[X_1^{(2)}]+\frac{3}{2}[M_1^{(4)}]-\frac{1}{2}[M_3^{(4)}]
\Big{)}
\notag \\
&\ \quad\quad\quad\quad\quad\quad\quad\quad\quad\quad\quad\quad\quad\quad\ (\text{mod}\ e).
\end{align}

\subsection{$C_6$-symmetry}
\begin{table}
\begin{tabular}{ccccc}
		\hline
		\hline
Label & Wannier configurations & \multicolumn{3}{c}{invariants}\\
		\hline
		\hline
	$j$ &	   & $[M^{(2)}_1]$ & $[K^{(3)}_1]$ & $[K^{(3)}_2]$ \\
		\hline
1&$W_{2b|_{0}}^{(6)}$ & 0 & $-2$ & 1 \\
2&$W_{2b|_{2\pi/3}}^{(6)}$ & 0 & 1 & 1 \\
3&$W_{2b|_{-2\pi/3}}^{(6)}$ & 0 & 1 & $-2$ \\
4&$W_{3c|_{0}}^{(6)}$ & $-2$ & 0 & 0 \\
5&$W_{3c|_{\pi}}^{(6)}$ & 2 & 0 & 0 \\
	\hline
	\hline
\end{tabular}
\caption{
Wannier configurations with labels and their invariants with $C_6$ symmetry in class A systems. The first column is the label for the WCs. 
The Wannier configurations belonging to Wyckoff positions $1a$ and $6d$ are omitted in the table because all topological invariants are 0.
}
	\label{tab:PrimitiveGenerators_C6}
\end{table}

In $C_6$-symmetric systems, we need to consider five Wannier configurations in Table~\ref{tab:PrimitiveGenerators_C6}. 
Table~\ref{tab:PrimitiveGenerators_C6} is summarized as follows:
\begin{align}
\begin{pmatrix}
[M^{(2)}_1] \\
[K^{(3)}_1] \\
[K^{(3)}_2] 
\end{pmatrix}
=
\begin{pmatrix}
0 & 0 & 0 & -2 & 2 \\
-2 & 1 & 1 & 0 & 0 \\
1 & 1 & -2 & 0 & 0 
\end{pmatrix}
\begin{pmatrix}
\alpha_1 \\
\alpha_2 \\
\vdots \\
\alpha_5 
\end{pmatrix}. 
\label{C6_WC_invariants}
\end{align}
Here, $M=\frac{1}{2}\bm{b}_1$ and $K=\frac{1}{3}\bm{b}_1+\frac{1}{3}\bm{b}_2$ (see Fig.~{\ref{Cn_BZ}}(c)). 
From the definition of $\alpha_j$, $n^{(e)}_{2b}=\alpha_1+\alpha_2+\alpha_3$ and $n^{(e)}_{3c}=\alpha_4+\alpha_5$. 
From Eq.~(\ref{C6_WC_invariants}), the following relation holds:
\begin{align}
[K_1^{(3)}]
&=-2\alpha_1+\alpha_2+\alpha_3
\notag \\
&\equiv \alpha_1+\alpha_2+\alpha_3\ (\text{mod}\ 3),
\\
\frac{1}{2}[M_1^{(2)}]
&=-\alpha_5+\alpha_6
\notag \\
&\equiv \alpha_5+\alpha_6\ (\text{mod}\ 2). 
\end{align}
Therefore, $n^{(e)}_{2b}$ (mod 3) and $n^{(e)}_{3c}$ (mod 2) can be determined from $[\Pi_{p}^{(n)}]$ as follows:
\begin{align}
n^{(e)}_{2b}&=[K_1^{(3)}]\ (\text{mod}\ 3), 
\\
n^{(e)}_{3c}&=\frac{1}{2}[M_1^{(2)}]\ (\text{mod}\ 2). 
\end{align}
Then, $n^{(e)}_{1a}=\nu-2n^{(e)}_{2b}-3n^{(e)}_{3c}-6n^{(e)}_{6d}\equiv\nu-2n^{(e)}_{2b}-3n^{(e)}_{3c}$ (mod 6) is also determined from $\nu$ and $[\Pi_{p}^{(n)}]$. 
Finally, the corner charge formula is summarized as follows: 
\begin{align}
Q_{c\_1a}^{(6)}
&\equiv\frac{|e|}{6}\Big{(}n_{1a}^{(\text{ion})}-\nu+2[K_1^{(3)}]+\frac{3}{2}[M_1^{(2)}]
\Big{)} \ (\text{mod}\ e). 
\end{align}

\section{Relationship between the topological invariants $[\Pi_{p}^{(n)}]$ and the $n^{(e)}_{mX}$ in class AI systems}
In class AI systems, as compared to class A systems, several Wannier configurations are connected by the time-reversal symmetry $\Theta$ ($\Theta^2=1$). Then, the Wannier configurations are changed as shown in Tables~\ref{tab:PrimitiveGenerators_C3_AI}, \ref{tab:PrimitiveGenerators_C4_AI} and 
\ref{tab:PrimitiveGenerators_C6_AI} for $C_3$, $C_4$ and $C_6$ symmetric systems, respectively.

\subsection{$C_3$-symmetry}
\begin{table}
\begin{tabular}{cccccc}
		\hline
		\hline
Label & Wannier configurations & \multicolumn{4}{c}{invariants}\\
		\hline
		\hline
	$j$ &	   & $[K^{(3)}_1]$ & $[K^{(3)}_2]$ & $[K^{\prime(3)}_1]$ & $[K^{\prime(3)}_2]$\\
		\hline
1&$W_{1b|_{0}}^{(3)}$ & $-1$ & 0 & $-1$ & 1 \\
2&$W_{1b|_{2\pi/3}}^{(3)}\oplus W_{1b|_{4\pi/3}}^{(3)}$ & 1 & 0 & 1 & $-1$ \\
3&$W_{1c|_{0}}^{(3)}$ & $-1$ & 1 & $-1$ & 0 \\
4&$W_{1c|_{2\pi/3}}^{(3)} \oplus W_{1c|_{4\pi/3}}^{(3)}$ & 1 & $-1$ & 1 & 0 \\
	\hline
	\hline
\end{tabular}
\caption{
Wannier configurations with labels and their invariants with $C_3$ symmetry in class AI systems. 
The first column is the label for the WCs. 
The Wannier configurations belonging to Wyckoff positions $1a$ and $3d$ are omitted in the table because all topological invariants are 0.}
	\label{tab:PrimitiveGenerators_C3_AI}
\end{table}

In $C_3$-symmetric systems, we need to consider four sets of Wannier configurations in Table~\ref{tab:PrimitiveGenerators_C3_AI}. 
Table~\ref{tab:PrimitiveGenerators_C3_AI} is summarized as follows:
\begin{align}
\begin{pmatrix}
[K^{(3)}_1] \\
[K^{(3)}_2] \\
[K^{\prime(3)}_1] \\
[K^{\prime(3)}_2] 
\end{pmatrix}
=
\begin{pmatrix}
-1 & 1 & -1 & 1 \\
0 & 0 & 1 & -1 \\
-1 & 1 & -1 & 1 \\
1 & -1 & 0 & 0 
\end{pmatrix}
\begin{pmatrix}
\alpha_1 \\
\alpha_2 \\
\alpha_3 \\
\alpha_4
\end{pmatrix}. 
\label{C3_WC_invariants_AI}
\end{align}
From the definition, $n^{(e)}_{1b}=\alpha_1\!+\!2\alpha_2$ and $n^{(e)}_{1c}=\alpha_3\!+\!2\alpha_4$. 
From Eq.~(\ref{C3_WC_invariants_AI}), the following relations hold:
\begin{align}
[K^{(3)}_1]&=[K^{\prime(3)}_1], 
\\
[K^{(3)}_2]+[K^{\prime(3)}_2]&=-[K^{(3)}_1], 
\label{KK_K}
\\
[K^{\prime(3)}_2]
&=\alpha_1-\alpha_2
\notag \\
&\equiv \alpha_1+2\alpha_2\ (\text{mod}\ 3),
\\
[K^{(3)}_2] 
&=\alpha_3-\alpha_4
\notag \\
&\equiv \alpha_3+2\alpha_4\ (\text{mod}\ 3). 
\end{align}
Therefore, $n^{(e)}_{1b}$ and $n^{(e)}_{1c}$ (mod $3$) can be determined from $\bm{\chi}$ as follows:
\begin{align}
&n^{(e)}_{1b}=[K^{\prime(3)}_2]
\ (\text{mod}\ 3), 
\\
&n^{(e)}_{1c}=[K^{(3)}_2]
\ (\text{mod}\ 3). 
\end{align}
Then, $n^{(e)}_{1a}=\nu-n^{(e)}_{1b}-n^{(e)}_{1c}-3n^{(e)}_{3d}\equiv\nu-([K^{(3)}_2]+[K^{\prime(3)}_2])\equiv\nu+[K^{(3)}_1]$ (mod 3). 
The corner charge formulas are represented as follows: 
\begin{align}
Q_{c\_1a}^{(3)}
&\equiv\frac{|e|}{3}\Big{(}n_{1a}^{(\text{ion})}-\nu-[K^{(3)}_1]\Big{)}\ (\text{mod}\ e),
\label{Qc3_AI_1a_Appendix}
\\
Q_{c\_1b}^{(3)}
&\equiv\frac{|e|}{3}\Big{(}n_{1b}^{(\text{ion})}-[K^{\prime(3)}_2]\Big{)}\ (\text{mod}\ e),
\\
Q_{c\_1c}^{(3)}
&\equiv\frac{|e|}{3}\Big{(}n_{1c}^{(\text{ion})}-[K^{(3)}_2]\Big{)}\ (\text{mod}\ e).
\end{align}

Our result in Eq.~(\ref{Qc3_AI_1a_Appendix}) is identical with Eq.~(\ref{Qc3_AI_1a}) in the main text. In particular, Eq.~(\ref{Qc3_AI_1a}) (Eq.~(\ref{Qc3_AI_1a_Appendix})) is reduced to the result of Ref.~[\onlinecite{PhysRevB.99.245151}] (i.e., Eq.~(\ref{Qc3_AI_1a_Benalcazar}) in the main text) provided (i) all the ions are located at $1a$. i.e., $n_{1a}^{(\text{ion})}=\nu$, and (ii) the bulk electric polarization is zero. 
Under these assumptions, from Eq.~(\ref{DeltaQ_3_1}) and $n_{1b}^{(\text{ion})}=n_{1c}^{(\text{ion})}=0$, 
\begin{align}
q_{b}+2q_{c}=-|e|(n^{(e)}_{1b}+2n^{(e)}_{1c})\equiv0\ (\text{mod}\ 3e). 
\end{align}
This means that $n^{(e)}_{1b}\equiv n^{(e)}_{1c}$ (mod 3), i.e., 
\begin{align}
[K^{\prime(3)}_2]\equiv[K^{(3)}_2]\ (\text{mod}\ 3). 
\end{align}
Then, from Eq.~(\ref{KK_K}), 
\begin{align}
[K^{(3)}_2]\equiv[K^{(3)}_1]\ (\text{mod}\ 3). 
\end{align}
Therefore, Eq.~(\ref{Qc3_AI_1a}) is reduced to Eq.~(\ref{Qc3_AI_1a_Benalcazar}) under the assumption that $n_{1a}^{(\text{ion})}=\nu$ and vanishing polarization.

\subsection{$C_4$-symmetry}
\begin{table}
\begin{tabular}{cccccc}\hline\hline
Label & Wannier configurations & \multicolumn{4}{c}{invariants}\\
		\hline
		\hline
	$j$	   & & $[X^{(2)}_1]$ & $[M^{(4)}_1]$ & $[M^{(4)}_2]$ & $[M^{(4)}_3]$\\
		\hline
1&$W_{1b|_{0}}^{(4)}$ & $-1$ & $-1$ & 0 & 1 \\
2&$W_{1b|_{\pi}}^{(4)}$ & $-1$ & 1 & 0 & $-1$ \\
3&$W_{1b|_{\pi/2}}^{(4)}\oplus W_{1b|_{-\pi/2}}^{(4)}$ & 2 & 0 & 0 & 0 \\
4&$W_{2c|_{0}}^{(4)}$ & $-1$ & $-1$ & 1 & $-1$ \\
5&$W_{2c|_{\pi}}^{(4)}$ & 1 & 1 & $-1$ & 1 \\
	\hline
	\hline
\end{tabular}
\caption{
Wannier configurations with labels and their invariants with $C_4$ symmetry in class AI systems. 
The first column is the label for the WCs. 
The Wannier configurations belonging to Wyckoff positions $1a$ and $4d$ are omitted in the table because all topological invariants are 0.}
	\label{tab:PrimitiveGenerators_C4_AI}
\end{table}

In $C_4$-symmetric systems, we need to consider five sets of Wannier configurations in Table~\ref{tab:PrimitiveGenerators_C4_AI}. 
Table~\ref{tab:PrimitiveGenerators_C4_AI} is summarized as follows:
\begin{align}
\begin{pmatrix}
[X^{(2)}_1] \\
[M^{(4)}_1] \\
[M^{(4)}_2] \\
[M^{(4)}_3] 
\end{pmatrix}
=
\begin{pmatrix}
-1 & -1 & 2 & -1 & 1 \\
-1 & 1 & 0 & -1 & 1 \\
0 & 0 & 0 & 1 & -1 \\
1 & -1 & 0 & -1 & 1 
\end{pmatrix}
\begin{pmatrix}
\alpha_1 \\
\alpha_2 \\
\vdots \\
\alpha_5 
\end{pmatrix}. 
\label{C4_WC_invariants_AI}
\end{align}
From the definition, $n^{(e)}_{1b}=\alpha_1+\alpha_2+2\alpha_3$ and $n^{(e)}_{2c}=\alpha_4+\alpha_5$. 
From Eq.~(\ref{C4_WC_invariants_AI}), the following relation holds:
\begin{align}
[X^{(2)}_1]-2[M^{(4)}_1]-[M^{(4)}_2]
&=\alpha_1-3\alpha_2+2\alpha_3
\notag \\
&\equiv \alpha_1+\alpha_2+2\alpha_3 \ (\text{mod}\ 4),
\\
[M^{(4)}_2]
&=\alpha_4-\alpha_5
\notag \\
&\equiv \alpha_4+\alpha_5 \ (\text{mod}\ 2).
\end{align}
Therefore, $n^{(e)}_{1b}$ (mod $4$) and $n^{(e)}_{1c}$ (mod $2$) can be determined from $[\Pi_{p}^{(n)}]$ as follows:
\begin{align}
&n^{(e)}_{1b}=[X^{(2)}_1]-2[M^{(4)}_1]-[M^{(4)}_2]\ (\text{mod}\ 4), 
\\
&n^{(e)}_{2c}
=[M_2^{(4)}]
\ (\text{mod}\ 2). 
\end{align}
Then, $n^{(e)}_{1a}=\nu-n^{(e)}_{1b}-2n^{(e)}_{2c}-4n^{(e)}_{4d}\equiv\nu-n^{(e)}_{1b}-2n^{(e)}_{2c}$ (mod 4) is also determined from $\nu$ and $[\Pi_{p}^{(n)}]$. 
Finally, the corner charge formulas are represented as follows: 
\begin{align}
Q_{c\_1a}^{(4)}
&\equiv\frac{|e|}{4}\Big{(}n_{1a}^{(\text{ion})}-\nu+[X_1^{(2)}]-2[M_1^{(4)}]+[M_2^{(4)}]
\Big{)} 
\notag \\
&\ \quad\quad\quad\quad\quad\quad\quad\quad\quad\quad\quad\quad\quad\quad\ (\text{mod}\ e),
\label{App_Q_1a_4_AI}
\\
Q_{c\_1b}^{(4)}
&\equiv\frac{|e|}{4}\Big{(}n_{1b}^{(\text{ion})}-[X^{(2)}_1]+2[M^{(4)}_1]+[M^{(4)}_2]
\Big{)}
\notag \\
&\ \quad\quad\quad\quad\quad\quad\quad\quad\quad\quad\quad\quad\quad\quad\ (\text{mod}\ e).
\end{align}

This Eq.~(\ref{App_Q_1a_4_AI}) is the same as Eq.~(\ref{Qc4_AI_1a}) in the main text. 
In particular, when $n_{1a}^{(\text{ion})}=\nu$ and vanishing polarization, Eq.~(\ref{Qc4_AI_1a}) is reduced to Eq.~(\ref{Qc4_AI_1a_Benalcazar}). 
Under these assumptions, from Eq.~(\ref{DeltaQ}) and $n_{1b}^{(\text{ion})}=n_{2c}^{(\text{ion})}=0$, 
\begin{align}
q_{b}+q_{c}=-|e|(n^{(e)}_{1b}+n^{(e)}_{2c})\equiv0\ (\text{mod}\ 2e). 
\end{align}
This means that $n^{(e)}_{1b}+n^{(e)}_{2c}\equiv0$ (mod 2), i.e., 
\begin{align}
[X^{(2)}_1]-2[M^{(4)}_1]\equiv[X^{(2)}_1]\equiv0
\ (\text{mod}\ 2). 
\end{align}
Therefore, Eq.~(\ref{Qc4_AI_1a}) is reduced to Eq.~(\ref{Qc4_AI_1a_Benalcazar}) under the assumption that $n_{1a}^{(\text{ion})}=\nu$ and vanishing polarization.

\subsection{$C_6$-symmetry}
\begin{table}
\begin{tabular}{ccccc}
		\hline
		\hline
Label & Wannier configurations & \multicolumn{3}{c}{invariants}\\
		\hline
		\hline
	$j$ &	   & $[M^{(2)}_1]$ & $[K^{(3)}_1]$ & $[K^{(3)}_2]$ \\
		\hline
1&$W_{2b|_{0}}^{(6)}$ & 0 & $-2$ & 1 \\
2&$W_{2b|_{2\pi/3}}^{(6)}\oplus W_{2b|_{-2\pi/3}}^{(6)}$ & 0 & 2 & $-1$ \\
3&$W_{3c|_{0}}^{(6)}$ & $-2$ & 0 & 0 \\
4&$W_{3c|_{\pi}}^{(6)}$ & 2 & 0 & 0 \\
	\hline
	\hline
\end{tabular}
\caption{
Wannier configurations with labels and their invariants with $C_6$ symmetry in class AI systems. The first column is the label for the WCs. 
The Wannier configurations belonging to Wyckoff positions $1a$ and $6d$ are omitted in the table because all topological invariants are 0.
}
	\label{tab:PrimitiveGenerators_C6_AI}
\end{table}

In $C_6$-symmetric systems, we need to consider four sets of Wannier configurations in Table~\ref{tab:PrimitiveGenerators_C6_AI}. 
Table~\ref{tab:PrimitiveGenerators_C6_AI} is summarized as follows:
\begin{align}
\begin{pmatrix}
[M^{(2)}_1] \\
[K^{(3)}_1] \\
[K^{(3)}_2] 
\end{pmatrix}
=
\begin{pmatrix}
0 & 0 & -2 & 2 \\
-2 & 2 & 0 & 0 \\
1 & -1 & 0 & 0 
\end{pmatrix}
\begin{pmatrix}
\alpha_1 \\
\alpha_2 \\
\alpha_3 \\
\alpha_4 
\end{pmatrix}. 
\label{C6_WC_invariants_AI}
\end{align}
Here, we briefly explain why $[K^{(3)}_1]\equiv[K^{(3)}_2]$ (mod 3). Due to the time-reversal symmetry, $[K^{(3)}_2]=[K^{(3)}_3]$. From the definition, $[K^{(3)}_1]+[K^{(3)}_2]+[K^{(3)}_3]=0$. Therefore, $[K^{(3)}_1]=-2[K^{(3)}_2]\equiv[K^{(3)}_2]$ (mod 3). 

From the definition, $n^{(e)}_{2b}=\alpha_1+2\alpha_2$ and $n^{(e)}_{3c}=\alpha_3+\alpha_4$. 
From Eq.~(\ref{C6_WC_invariants_AI}), the following relations hold:
\begin{align}
[K_1^{(3)}]
&=-2\alpha_1+2\alpha_2
\notag \\
&\equiv \alpha_1+2\alpha_2\ (\text{mod}\ 3),
\\
\frac{1}{2}[M_1^{(2)}]
&=-\alpha_3+\alpha_4
\notag \\
&\equiv \alpha_3+\alpha_4\ (\text{mod}\ 2). 
\end{align}
Therefore, $n^{(e)}_{2b}$ (mod 3) and $n^{(e)}_{3c}$ (mod 2) can be determined from $[\Pi_{p}^{(n)}]$ as follows:
\begin{align}
n^{(e)}_{2b}&=[K_1^{(3)}]\ (\text{mod}\ 3), 
\\
n^{(e)}_{3c}&=\frac{1}{2}[M_1^{(2)}]\ (\text{mod}\ 2). 
\end{align}
Then, $n^{(e)}_{1a}=\nu-2n^{(e)}_{2b}-3n^{(e)}_{3c}-6n^{(e)}_{6d}\equiv\nu-2n^{(e)}_{2b}-3n^{(e)}_{3c}$ (mod 6) is also determined from $\nu$ and $[\Pi_{p}^{(n)}]$. 
Finally, the corner charge formula is summarized as follows: 
\begin{align}
Q_{c\_1a}^{(6)}
&\equiv\frac{|e|}{6}\Big{(}n_{1a}^{(\text{ion})}-\nu+2[K_1^{(3)}]+\frac{3}{2}[M_1^{(2)}]
\Big{)} \ (\text{mod}\ e),
\end{align}
which is the same as Eq.~(\ref{Qc6_AI_1a}) in the main text.

Because both $[M^{(2)}_1]$ and $[K^{(3)}_1]$ are even numbers from Eq.~(\ref{C6_WC_invariants_AI}), Eq.~(\ref{Qc6_AI_1a}) is equivalent to Eq.~(\ref{Qc6_AI_1a_Benalcazar}), provided $n^{(\text{ion})}_{1a}=\nu$.

\section{Relationship between the topological invariants $[\Pi_{p}^{(n)}]$ and the $n^{(e)}_{mX}$ in class AII systems}
In class AII systems, several Wannier configurations are connected by the time-reversal symmetry $\Theta$ ($\Theta^2=-1$). Wannier configurations acquire double degeneracy due to the Kramers theorem. Then, the Wannier configurations are changed as shown in Tables~\ref{tab:PrimitiveGenerators_C3_AII}, \ref{tab:PrimitiveGenerators_C4_AII}, \ref{tab:PrimitiveGenerators_C6_AII}.

\begin{table}
\begin{tabular}{cccccc}
		\hline
		\hline
Label & Wannier configurations & \multicolumn{4}{c}{invariants}\\
		\hline
		\hline
	$j$ &	   & $[K^{(3)}_1]$ & $[K^{(3)}_2]$ & $[K^{\prime(3)}_1]$ & $[K^{\prime(3)}_2]$\\
		\hline
1&$2\times W_{1b|_{\pi}}^{(3)}$ & 2 & $-2$ & 0 & $-2$ \\
2&$W_{1b|_{\pi/3}}^{(3)}\oplus W_{1b|_{-\pi/3}}^{(3)}$ & $-1$ & 1 & 0 & 1 \\
3&$2\times W_{1c|_{\pi}}^{(3)}$ & 0 & $-2$ & 2 & $-2$ \\
4&$W_{1c|_{\pi/3}}^{(3)} \oplus W_{1c|_{-\pi/3}}^{(3)}$ & 0 & 1 & $-1$ & 1 \\
	\hline
	\hline
\end{tabular}
\caption{
Wannier configurations with labels and their invariants with $C_3$ symmetry in class AII systems. 
The first column is the label for the WCs. 
The Wannier configurations belonging to Wyckoff positions $1a$ and $3d$ are omitted in the table because all topological invariants are 0.}
	\label{tab:PrimitiveGenerators_C3_AII}
\end{table}

In $C_3$-symmetric systems, we need to consider four sets of Wannier configurations in Table~\ref{tab:PrimitiveGenerators_C3_AII}. 
Table~\ref{tab:PrimitiveGenerators_C3_AII} is summarized as follows:
\begin{align}
\begin{pmatrix}
[K^{(3)}_1] \\
[K^{(3)}_2] \\
[K^{\prime(3)}_1] \\
[K^{\prime(3)}_2] 
\end{pmatrix}
=
\begin{pmatrix}
2 & -1 & 0 & 0 \\
-2 & 1 & -2 & 1 \\
0 & 0 & 2 & -1 \\
-2 & 1 & -2 & 1  
\end{pmatrix}
\begin{pmatrix}
\alpha_1 \\
\alpha_2 \\
\alpha_3 \\
\alpha_4
\end{pmatrix}. 
\label{C3_WC_invariants_AII}
\end{align}
From the definition, $n^{(e)}_{1b}=2(\alpha_1\!+\!\alpha_2)$ and $n^{(e)}_{1c}=2(\alpha_3\!+\!\alpha_4)$. 
From Eq.~(\ref{C3_WC_invariants_AII}), the following relations hold:
\begin{align}
[K^{(3)}_2]&=[K^{\prime(3)}_2], 
\\
[K^{(3)}_1]+[K^{\prime(3)}_1]&=-[K^{(3)}_2], 
\\
[K^{(3)}_1]
&=2\alpha_1-\alpha_2
\notag \\
&\equiv 2(\alpha_1+\alpha_2)\ (\text{mod}\ 3),
\\
[K^{\prime(3)}_1] 
&=2\alpha_3-\alpha_4
\notag \\
&\equiv 2(\alpha_3+\alpha_4)\ (\text{mod}\ 3). 
\end{align}
Therefore, $n^{(e)}_{1b}$ and $n^{(e)}_{1c}$ (mod $3$) can be determined from $\bm{\chi}$ as follows:
\begin{align}
&n^{(e)}_{1b}=[K^{(3)}_1]
\ (\text{mod}\ 3), 
\\
&n^{(e)}_{1c}=[K^{\prime(3)}_1]
\ (\text{mod}\ 3). 
\end{align}
Then, $n^{(e)}_{1a}=\nu-n^{(e)}_{1b}-n^{(e)}_{1c}-3n^{(e)}_{3d}\equiv\nu-([K^{(3)}_1]+[K^{\prime(3)}_1])\equiv\nu+[K^{(3)}_2]$ (mod 3). 
The corner charge formulas are represented as follows: 
\begin{align}
Q_{c\_1a}^{(3)}
&\equiv\frac{|e|}{3}\Big{(}n_{1a}^{(\text{ion})}-\nu-[K^{(3)}_2]\Big{)}\ (\text{mod}\ e),
\label{Qc3_AII_1a_Appendix}
\\
Q_{c\_1b}^{(3)}
&\equiv\frac{|e|}{3}\Big{(}n_{1b}^{(\text{ion})}-[K^{(3)}_1]\Big{)}\ (\text{mod}\ e),
\\
Q_{c\_1c}^{(3)}
&\equiv\frac{|e|}{3}\Big{(}n_{1c}^{(\text{ion})}-[K^{\prime(3)}_1]\Big{)}\ (\text{mod}\ e).
\end{align}

\begin{table}
\begin{tabular}{cccccc}\hline\hline
Label & Wannier configurations & \multicolumn{4}{c}{invariants}\\
		\hline
		\hline
	$j$	   & & $[X^{(2)}_1]$ & $[M^{(4)}_1]$ & $[M^{(4)}_2]$ & $[M^{(4)}_3]$\\
		\hline
1&$W_{1b|_{\pi/4}}^{(4)}\oplus W_{1b|_{-\pi/4}}^{(4)}$ & 0 & $-1$ & 1 & 1 \\
2&$W_{1b|_{3\pi/4}}^{(4)}\oplus W_{1b|_{-3\pi/4}}^{(4)}$ & 0 & 1 & $-1$ & $-1$ \\
	\hline
	\hline
\end{tabular}
\caption{
Wannier configurations with labels and their invariants with $C_4$ symmetry in class AII systems. 
The first column is the label for the WCs. 
The Wannier configurations belonging to Wyckoff positions $1a$, $2c$ and $4d$ are omitted in the table because all topological invariants are 0.}
	\label{tab:PrimitiveGenerators_C4_AII}
\end{table}

In $C_4$-symmetric systems, we need to consider two sets of Wannier configurations in Table~\ref{tab:PrimitiveGenerators_C4_AII}. 
Table~\ref{tab:PrimitiveGenerators_C4_AII} is summarized as follows:
\begin{align}
\begin{pmatrix}
[X^{(2)}_1] \\
[M^{(4)}_1] \\
[M^{(4)}_2] \\
[M^{(4)}_3] 
\end{pmatrix}
=
\begin{pmatrix}
0 & 0 \\
-1 & 1 \\
1 & -1 \\
1 & -1 
\end{pmatrix}
\begin{pmatrix}
\alpha_1 \\
\alpha_2 
\end{pmatrix}. 
\label{C4_WC_invariants_AII}
\end{align}
From the definition, $n^{(e)}_{1b}=2(\alpha_1\!+\!\alpha_2)$. 
From Eq.~(\ref{C4_WC_invariants_AII}), the following relations hold:
\begin{align}
2[M^{(4)}_1]=2(-\alpha_1+\alpha_2)&\equiv2(\alpha_1+\alpha_2)\ (\text{mod}\ 4),
\\
n^{(e)}_{1b}&\equiv2[M^{(4)}_1]\ (\text{mod}\ 4).
\end{align}
Then, $n^{(e)}_{1a}=\nu-n^{(e)}_{1b}-2n^{(e)}_{2c}-4n^{(e)}_{4d}\equiv\nu-2[M^{(4)}_1]$ (mod 4). Here, we used $n^{(e)}_{2c}\equiv0$ (mod 2) due to the Kramers degeneracy. 
The corner charge formulas are summarized as follows:
\begin{align}
Q_{c\_1a}^{(4)}
&\equiv\frac{|e|}{4}\Big{(}n_{1a}^{(\text{ion})}-\nu+2[M^{(4)}_1]\Big{)}\ (\text{mod}\ e),
\label{Qc4_AII_1a_Appendix}
\\
Q_{c\_1b}^{(4)}
&\equiv\frac{|e|}{4}\Big{(}n_{1b}^{(\text{ion})}-2[M^{(4)}_1]\Big{)}\ (\text{mod}\ e).
\end{align}

In $C_6$-symmetric systems, we need to consider two sets of Wannier configurations in Table~\ref{tab:PrimitiveGenerators_C6_AII}. 
Table~\ref{tab:PrimitiveGenerators_C6_AII} is summarized as follows:
\begin{align}
\begin{pmatrix}
[M^{(2)}_1] \\
[K^{(3)}_1] \\
[K^{(3)}_2] 
\end{pmatrix}
=
\begin{pmatrix}
0 & 0 \\
2 & -1 \\
-4 & 2 
\end{pmatrix}
\begin{pmatrix}
\alpha_1 \\
\alpha_2 
\end{pmatrix}. 
\label{C6_WC_invariants_AII}
\end{align}
From the definition, $n^{(e)}_{2b}=2(\alpha_1+\alpha_2)$. 
From Eq.~(\ref{C6_WC_invariants_AII}), the following relations hold:
\begin{align}
[K^{(3)}_1]=2\alpha_1-\alpha_2&\equiv2(\alpha_1+\alpha_2)\ (\text{mod}\ 3),
\\
n^{(e)}_{2b}&\equiv[K^{(3)}_1]\ (\text{mod}\ 3).
\end{align}
Then, $n^{(e)}_{1a}=\nu-2n^{(e)}_{2b}-3n^{(e)}_{3c}-6n^{(e)}_{6d}\equiv\nu-2[K^{(3)}_1]$ (mod 6). Here, we used $n^{(e)}_{3c}\equiv0$ (mod 2) due to the Kramers degeneracy. 
The corner charge formula is summarized as follows:
\begin{align}
Q_{c\_1a}^{(6)}
&\equiv\frac{|e|}{6}\Big{(}n_{1a}^{(\text{ion})}-\nu+2[K^{(3)}_1]\Big{)}\ (\text{mod}\ e).
\label{Qc6_AII_1a_Appendix}
\end{align}

\begin{table}
\begin{tabular}{ccccc}
		\hline
		\hline
Label & Wannier configurations & \multicolumn{3}{c}{invariants}\\
		\hline
		\hline
	$j$ &	   & $[M^{(2)}_1]$ & $[K^{(3)}_1]$ & $[K^{(3)}_2]$ \\
		\hline
1&$2\times W_{2b|_{\pi}}^{(6)}$ & 0 & 2 & $-4$ \\
2&$W_{2b|_{\pi/3}}^{(6)}\oplus W_{2b|_{-\pi/3}}^{(6)}$ & 0 & $-1$ & 2 \\
	\hline
	\hline
\end{tabular}
\caption{
Wannier configurations with labels and their invariants with $C_6$ symmetry in class AII systems. 
The first column is the label for the WCs. 
The Wannier configurations belonging to Wyckoff positions $1a$, $3c$ and $6d$ are omitted in the table because all topological invariants are 0.
}
	\label{tab:PrimitiveGenerators_C6_AII}
\end{table}


\end{document}